\documentclass[11pt,a4paper]{article}
\pdfoutput=1
\usepackage{jheppub} 
\usepackage{amsfonts,amssymb,latexsym,amsmath,mathrsfs,amsbsy}
\usepackage{bm}
\usepackage{slashed,bbold}
\usepackage[hhmmss]{datetime}
\usepackage{dsfont}
\usepackage{feynmf}

\usepackage[latin1]{inputenc}

\title{Torsional Constitutive Relations at Finite Temperature} 

\author[a]{Manuel Valle}
\author[b]{and Miguel \'A. V\'azquez-Mozo}

\affiliation[a]{Departamento de F\'\i sica, 
Universidad del Pa\'is Vasco UPV/EHU, \\
Apartado 644,  48080 Bilbao, Spain}
\affiliation[b]{Departamento de F\'\i sica Fundamental, Universidad de Salamanca, \\
Plaza de la Merced s/n, 37008 Salamanca, Spain}

\emailAdd{manuel.valle@ehu.es}
\emailAdd{vazquez@usal.es}

\abstract{
The general form of the linear torsional constitutive relations at finite temperature 
of the chiral current, energy-momentum tensor, and spin energy potential
are computed for a chiral fermion fluid minimally coupled to geometric torsion and with
nonzero chiral chemical potential. 
The corresponding transport coefficients are 
explicitly calculated in terms of the energy
and number densities evaluated at vanishing torsion.
A microscopic calculation of these constitutive relations in some
particular backgrounds is also presented, confirming the general structure found.
}

\arxivnumber{}

\begin{document}

\maketitle


\flushbottom



\section{Introduction}
\label{sec:intro}

Transport in the presence of a background geometric torsion~\cite{Hehl:1976kj,Shapiro:2001rz} 
has been the subject of recent interest, among other reasons because it
provides a way of incorporating the effect of lattice dislocations
on electronic fluids~\cite{Kondo1952,Bilby:1955,Katanaev:1992kh,Hehl:2007bn,Kleman:2008zz,Katanaev:2021bje}. 
Although much of the attention is focused on the role
of torsion in chiral nondissipative transport~\cite{Hughes:2012vg,Parrikar:2014usa,
Khaidukov:2018oat,Nissinen:2019kld,Huang:2019haq,Nissinen:2019wmh,Nissinen:2019mkw,
Huang:2019adx,Huang:2020ypv,Imaki:2020csc,
Ferreiros:2020uda,Manes:2020zdd,Liu:2021bic,Chernodub:2021nff,Nissinen:2021gke,Valle:2021nfv,Amitani:2022xev},
other aspects have also been widely
addressed in the literature~\cite{Hidaka:2012rj,Gromov:2014vla,Geracie:2014mta,
Sumiyoshi:2015eda,Laurila:2020yll,
Valle:2015hfa,Huang:2021luf,Nissinen:2023bgl},
particularly in the context of spin 
hydrodynamics~\cite{Gallegos:2020otk,Gallegos:2021bzp,Hongo:2021ona,Gallegos:2022jow} 
(see~\cite{Florkowski:2018fap,Bhadury:2021oat,Hattori:2022hyo} for reviews). 
The issue of the possible experimental detection of torsional transport phenomena 
has also been considered~\cite{Chu:2022bhj}. 

In ref.~\cite{Valle:2021nfv} we examined the linear torsional constitutive relations at
zero temperature for a fluid of chiral fermions minimally coupled to 
geometric torsion. 
A descent analysis was applied to construct the Chern-Simons equilibrium partition function
from the six-form torsional anomaly polynomial. 
This in turn was used to find the 
general form of the torsional constitutive relations, all of
them expressed in terms of the model- and cutoff-dependent global normalization of the anomaly polynomial. 

The aim of the present work is to go beyond vacuum contributions 
and study thermal linear torsional corrections to the 
constitutive relations of a chiral fermion fluid minimally coupled to torsion and in the presence of
a nonzero chiral chemical potential. Exploiting the equivalence 
with the coupling to an effective
axial gauge field~\cite{Hehl:1971qi,Datta:1971id}, we determine
the form of the equilibrium partition function linear in the torsion to first order in the 
derivative expansion. 
From this, the finite temperature constitutive relations for the
chiral current, energy-momentum tensor, and spin energy potential are obtained, 
finding closed expressions for
all of them. 
The corresponding transport coefficients are expressed in terms of thermodynamical
functions evaluated at vanishing torsion, in particular the number and energy densities, as well as the 
derivative of the latter with respect to the chiral chemical potential. 

We also present a first-principles
computation of the finite temperature constitutive relations for three particular torsional backgrounds by 
solving the corresponding thermal one-particle Green function equations at linear order in the
background data. An important feature
of all the three cases considered
is that they have nonzero spin connection and curvature.
Besides serving as a crosscheck of the general results obtained, these analyses also shed some light on
the interplay between torsion and other background data, most particularly vorticity. 
Another worth-mentioning result about the examples studied is that the torsional contributions to
the vacuum, zero-temperature parts of the chiral current can be
set to zero by an appropriate regularization prescription. The same happens with the 
Nieh-Yan anomaly, in the case when it receives contributions at linear order in torsion.

The paper is organized as follows. 
In section~\ref{sec:general_analysis} we present the calculation of the general structure of 
the finite temperature
linear torsional corrections to the chiral current, energy-momentum tensor, and spin 
energy potential of a chiral fermion fluid.  
Section~\ref{sec:microscopic} is devoted to the analysis of 
the finite temperature constitutive relations
of a chiral fluid 
in a background with purely magnetic torsion and nonvanishing curvature that 
generalizes the one used in refs.~\cite{Huang:2019haq,Laurila:2020yll} 
to describe torsional Landau levels. In section~\ref{sec:T0ij} a second geometry
is investigated, this time 
with magnetic torsion and vorticity, extending the background studied in
ref.~\cite{Khaidukov:2018oat}, where all constitutive relations
are free from torsional corrections at linear order. 
As a last instance, in sec.~\ref{sec:NY_anomaly} we consider a geometry describing a 
chiral fermion fluid with electric torsion and nonzero spin chemical potential for 
which the chiral current, unlike the other particular geometries considered earlier, is affected by 
a nonzero Nieh-Yan anomaly. 
Finally, our results are summarized and discussed 
in sec.~\ref{sec:conclusions}.

\section{Linear torsional constitutive relations at finite temperature}
\label{sec:general_analysis}

In order to study the general structure of the linear torsional thermal corrections 
for a fluid of chiral fermions,
we start from the action of a left-handed Weyl fermion 
minimally coupled to torsion and to an external gauge field~$\mathcal{A}_{\mu}$~\cite{Freedman:2012zz}
\begin{align}
S=-{i\over 2}\int d^{4}x\,(\det{e})\Big(\overline{\psi}\overline{\sigma}_{-}^{\mu}
\overrightarrow{\nabla}_{\mu}\psi
-\overline{\psi}\overleftarrow{\nabla}_{\mu}\overline{\sigma}^{\mu}_{-}\psi
-2i\psi^{\dagger}\overline{\sigma}_{-}^{\mu}\psi\mathcal{A}_{\mu}\Big),
\label{eqw:action_full}
\end{align}
where the covariant derivatives are defined in terms of the spin connection by
\begin{align}
\overrightarrow{\nabla}_{\!\!\mu}\psi&=\partial_{\mu}\psi+{1\over 4}
\sigma_{+}^{[a}\sigma_{-}^{b]}\omega_{ab\mu}\psi,\nonumber \\[0.2cm]
\psi^{\dagger}\overleftarrow{\nabla}_{\!\!\mu}&=\partial_{\mu}\psi^{\dagger}
-{1\over 4}\psi^{\dagger}\sigma_{-}^{[a}\sigma_{+}^{b]}\omega_{ab\mu}.
\label{eq:right_left_derivs}
\end{align}
In writing these expressions, we have introduced the 
matrices~$\overline{\sigma}^{\mu}_{\pm}\equiv e_{a}^{\,\,\,\mu}\sigma^{a}_{\pm}$, 
where~$e_{a}^{\,\,\,\mu}$ is the vierbein 
and~$\sigma_{\pm}^{a}$ are given in terms of the three Pauli matrices by
$\sigma_{\pm}^{a}\equiv(\pm\mathbb{1},\sigma_{x},\sigma_{y},\sigma_{z})$. 
We also assume nonvanishing geometric torsion\footnote{Greek letters are used throughout the 
paper for spacetime indices. Spatial coordinates are indicated by~$i,j,\ldots$, while~$a,b,\ldots$
denote four-dimensional Lorentz indices.  For simplicity, we
also omit the wedge symbol~$\wedge$ to indicate the exterior product.} 
\begin{align}
T^{a}=de^{a}+\omega^{a}_{\,\,\,\,b}e^{b},
\label{eq:torsion_two_form}
\end{align}
and write the 
spin connection as~$\omega^{a}_{\,\,\,b}
=\overline{\omega}^{a}_{\,\,\,b}+\kappa^{a}_{\,\,\,b}$, with~$\overline{\omega}^{a}_{\,\,\,b}$ 
the Levi-Civita part 
satisfying~$de^{a}+\overline{\omega}^{a}_{\,\,\,\,b}
e^{b}=0$ and~$\kappa^{a}_{\,\,\,b}\equiv \kappa^{a}_{\,\,\,b\mu}dx^{\mu}$ the contortion tensor~\cite{Hehl:1976kj,Freedman:2012zz}. 
Using a number of Pauli matrices identities,
the action~\eqref{eqw:action_full} can be recast as
\begin{align}
S&=-{i\over 2}\int d^{4}x\,(\det{e})\Big(\psi^{\dagger}\sigma^{\mu}_{-}
\overrightarrow{\overline{\nabla}}_{\!\!\mu}\psi
-\psi^{\dagger}\overleftarrow{\overline{\nabla}}_{\!\!\mu}\overline{\sigma}^{\mu}_{-}\psi
-2i\psi^{\dagger}\overline{\sigma}_{-}^{\mu}\psi\mathsf{A}_{\mu}\Big).
\label{eq:action_effective_axial}
\end{align}
The barred covariant derivatives are obtained by replacing the 
full spin connection~$\omega^{a}_{\,\,\,\,b}$ 
in~\eqref{eq:right_left_derivs} with its Levi-Civita 
part~$\overline{\omega}^{a}_{\,\,\,\,b}$, while
the effective gauge field~$\mathsf{A}_{\mu}$ is
defined by
\begin{align}
\mathsf{A}_{\mu}\equiv\mathcal{A}_{\mu}+\mathcal{S}_{\mu}.
\end{align}
The field 
\begin{align}
\mathcal{S}_{\mu}&=-{1\over 4}\epsilon_{\mu\nu}^{\,\,\,\,\,\,\,\alpha\beta}\kappa^{\nu}_{\,\,\,\,\alpha\beta}
\nonumber \\[0.2cm]
&={1\over 8}\epsilon_{\mu\nu}^{\,\,\,\,\,\,\,\alpha\beta}T^{\nu}_{\,\,\,\,\alpha\beta},
\label{eq:eff_gauge_field}
\end{align}
encodes all torsion dependence,
where the spacetime components of the torsion tensor are defined by
\begin{align}
T^{\mu}=e_{a}^{\,\,\,\,\mu}T^{a}={1\over 2}T^{\mu}_{\,\,\,\alpha\beta}dx^{\alpha}dx^{\beta},
\end{align}
and similarly for the ones of the 
contortion tensor,~$\kappa^{\mu}_{\,\,\alpha\beta}=e_{a}^{\,\,\,\mu}e^{b}_{\,\,\,\alpha}\kappa^{a}_{\,\,\,b\beta}$.
The effect of background torsion can be thus recast as a coupling of the chiral fermions to an effective external 
gauge field~\cite{Hehl:1971qi,Datta:1971id}.

It should be kept in mind that in all our expressions the four-dimensional
Levi-Civita tensor is normalized according to~$\epsilon_{0123}=\sqrt{-G}$ and~$\epsilon^{0123}
=-1/\sqrt{-G}$, with~$G$ the determinant of the spacetime metric. 
Notice that the expression of the torsional gauge field one-form involves a Hodge 
dual,~$\mathcal{S}\equiv \mathcal{S}_{\mu}dx^{\mu}=-{1\over 4}\star(e_{a}T^{a})$, and therefore its components
depend on the background metric. This fact will play an important role later on
in the analysis of the constitutive relations of the energy-momentum tensor components.

Using the identity~$\overline{\sigma}_{-}^{(\mu}\overline{\sigma}_{+}^{\nu)}=G^{\mu\nu}\mathbb{1}$, 
the Dirac-Weyl equation derived from~\eqref{eq:action_effective_axial}
can be recast as
\begin{align}
i\partial_{0}\psi=-\left({i\over 4}\sigma_{+}^{[a}\sigma_{-}^{b]}
\overline{\omega}_{ab0}+\mathsf{A}_{0}\right)\psi
-{1\over G^{00}}\overline{\sigma}_{+}^{0}\overline{\sigma}_{-}^{k}\big(i\overline{\nabla}_{k}
+\mathsf{A}_{k}\big)\psi.
\end{align}
From this, we read the one-particle Hamiltonian operator
\begin{align}
\mathcal{H}=-\left({i\over 4}\sigma_{+}^{[a}\sigma_{-}^{b]}\overline{\omega}_{ab0}
+\mathsf{A}_{0}\right)
-{1\over G^{00}}\overline{\sigma}_{+}^{0}\overline{\sigma}_{-}^{k}\big(i\overline{\nabla}_{k}
+\mathsf{A}_{k}\big),
\label{eq:hamiltonian_general}
\end{align}
which is self-adjoint with respect to the inner product
\begin{align}
\langle\psi_{1}|\psi_{2}\rangle\equiv 
-\int_{\Sigma_{t}} d^{3}x\,\sqrt{-G}\,\psi_{1}^{\dagger}
\overline{\sigma}_{+}^{0}
\psi_{2},
\end{align}
with~$\Sigma_{t}$ a constant-time section and~$G$ the metric determinant. 

\paragraph{The equilibrium partition function.}
To study a fluid of negative chirality fermions in the presence of geometric
torsion, we follow the effective action 
approach~\cite{Banerjee:2012iz,Jensen:2012jh}, assuming that
chiral fermions propagate 
on a generic static background spacetime metric
\begin{align}
ds^{2}\equiv G_{\mu\nu}dx^{\mu}dx^{\nu}
=-e^{2\sigma(\mathbf{x})}\big[dt+a_{i}(\mathbf{x})dx^{i}\big]^{2}+g_{ij}(\mathbf{x})dx^{i}dx^{j}.
\label{eq:static_metric}
\end{align} 
The fluid's equilibrium four-velocity, acceleration, and vorticity are related 
to the metric functions by the standard expressions
\begin{align}
u(\mathbf{x})&\equiv -e^{\sigma(\mathbf{x})}\big[dt+a_{i}(\mathbf{x})dx^{i}\big], \nonumber \\[0.2cm]
\mathfrak{a}(\mathbf{x})&=d\sigma(\mathbf{x}), \label{eq:uaomega}\\[0.2cm]
\omega(\mathbf{x})&=-{1\over 2}e^{\sigma(\mathbf{x})}da(\mathbf{x}),
\nonumber
\end{align}
while the chiral chemical potential and 
the local equilibrium temperature are respectively defined
by
\begin{align}
\mu_{L}(\mathbf{x})&=e^{-\sigma(\mathbf{x})}\mathcal{A}_{0},\nonumber \\[0.2cm] 
T(\mathbf{x})&=e^{-\sigma(\mathbf{x})}T_{0},
\label{eq:muLT0defs}
\end{align} 
with~$T_{0}^{-1}\equiv \beta_{0}$ the length of the compatified thermal Euclidean circle
in the imaginary-time formalism to be used in our computations of the following sections. 
In the absence of torsion, the equilibrium partition function can be written
as\footnote{Here and in the rest of the paper 
superscripts inside parentheses are used to 
indicate the order in the torsion or, generically, the background data
of the corresponding quantity.}
\begin{align}
W^{(0)}&={1\over T_{0}}\int d^{3}x\sqrt{g}\,e^{\sigma}\mathcal{P}(T,\mu_{L}),
\label{eq:W0der}
\end{align}
where~$\mathcal{P}(T,\mu_{L})$ is the fluid's 
pressure,~$\mu_{L}=u^{\mu}\mathcal{A}_{\mu}$ the chiral chemical potential,
and~$g$ the determinant of the
transverse metric in~\eqref{eq:static_metric}. It should be taken into account that gauge invariance
prevents the pressure from depending on the spatial components of the gauge field.
Although a dependence on the static electric and magnetic fields is allowed in principle, it would induce 
linear torsional corrections only at second order in derivatives. This is the reason why we
do not consider this possibility here. 

From the form of the microscopic action~\eqref{eq:action_effective_axial}, 
we reach the key observation that the coupling to 
background torsion is implemented in the equilibrium partition function by a shift of the chiral
chemical potential
\begin{align}
\mu_{L}&\longrightarrow \mu_{L}
+\mu_{S},
\label{eq:shift}
\end{align} 
where~$\mu_{S}\equiv u^{\mu}\mathcal{S}_{\mu}$ contains all dependence on the torsion 
tensor~\cite{Manes:2020zdd}
\begin{align}
\mu_{S}&={1\over 8}\epsilon^{\mu\nu\alpha\beta}u_{\mu}T_{\nu\alpha\beta}
\nonumber \\[0.2cm]
&=-{1\over 8}\epsilon^{ijk}g_{i\ell}\big(T^{\ell}_{\,\,\,jk}-a_{j}T^{\ell}_{\,\,\,0k}-a_{k}T^{\ell}_{\,\,\,j0}\big).
\label{eq:muS_spatial_comp}
\end{align}
To write the expression in the second line, we used the explicit form of 
the fluid four-velocity in~\eqref{eq:uaomega},
taking into account the 
antisymmetry of~$T^{\mu}_{\,\,\,\nu\alpha}$ in its two lower indices, as well as the definition of the  
three-dimensional Levi-Civita tensor~$\epsilon^{ijk}=-e^{\sigma}\epsilon^{0ijk}$, 
normalized as~$\epsilon^{123}=1/\sqrt{g}$.
Equation~\eqref{eq:muS_spatial_comp} explicitly shows that~$\mu_{S}$ is independent of~$\sigma$
and also invariant
under the Kaluza-Klein (KK) transformations induced by 
space-dependent time reparametrizations,~$t\rightarrow
t+\phi(\mathbf{x})$.

Following ref.~\cite{Gallegos:2022jow}, the contortion tensor is taken to be first order in the 
derivative expansion\footnote{Besides the compelling arguments provided in this reference, classifying the contortion tensor as first order
in derivatives is also suggested by the identification
of torsion and vorticity in certain backgrounds~\cite{Khaidukov:2018oat}.}.
This means that, unlike the standard bona fide microscopic 
gauge field~$\mathcal{A}_{\mu}$, the effective gauge field~\eqref{eq:eff_gauge_field}
is first order in derivatives, an important fact to bear in mind. 
Thus, implementing the shift~$\mu_{L}\rightarrow\mu_{L}+\mu_{S}$ in eq.~\eqref{eq:W0der} and expanding
to linear order in~$\mu_{S}$, we obtain the linear torsional correction to the equilibrium partition function 
at first order in derivatives
\begin{align}
W^{(1)}&={1\over T_{0}}\int d^{3}x\sqrt{g}\,e^{\sigma}\langle n\rangle \mu_{S} \nonumber \\[0.2cm]
&=-{1\over 8T_{0}}\int d^{3}x\sqrt{g}\,e^{\sigma}\langle n\rangle \epsilon^{\mu\nu\alpha\beta}u_{\mu}
T^{\sigma}_{\,\,\,\alpha\beta}G_{\nu\sigma},
\label{eq:W(1)}
\end{align}
where~$\langle n\rangle$ is the thermal-averaged number density
\begin{align}
\langle n \rangle\equiv\left. {\partial \mathcal{P}\over \partial \mu_{L}}\right|_{\mu_{S}=0},
\label{eq:<n>_def}
\end{align}
and the subscript on the right-hand side indicates that the derivative is evaluated at vanishing torsion. 

\paragraph{Chiral current.}
The linear torsional corrections to the chiral current are found applying the expression given  
ref.~\cite{Banerjee:2012iz} to the partition function~\eqref{eq:W(1)}, taking into account
the relation between~$\mathcal{A}_{0}$ and~$\mu_{L}$ shown in eq.~\eqref{eq:muLT0defs}.
For the (covariant) zero component, we obtain
\begin{align}
\langle J_{0}\rangle^{(1)}
&\equiv -{T_{0}\over \sqrt{g}}{\delta W^{(1)}\over \delta \mu_{L}} \nonumber \\[0.2cm]
&=-\mu_{S}e^{\sigma}{\partial\langle n\rangle\over \partial\mu_{L}},
\label{eq:J_0_general}
\end{align}
while the (contravariant) spatial components are equal to zero
\begin{align}
\langle J^{i}\rangle^{(1)}
&\equiv{T_{0}e^{-\sigma}\over\sqrt{g}}{\delta W^{(1)}\over \delta A_{i}}=0.
\label{eq:J^i_general}
\end{align}
This last result follows from the independence of the pressure with respect to~$A_{i}$, as explained after eq.~\eqref{eq:W0der}.

\paragraph{The energy-momentum tensor.}
Besides parametrizing velocity, vorticity, and acceleration in the fluid, the metric
functions in~\eqref{eq:static_metric} are the classical sources coupling to the
components of the fluid energy-momentum tensor, which are computed by taking
functional derivatives of the 
partition function with respect to them~\cite{Banerjee:2012iz}. 
Let us recall that, at the level of the microscopic action~\eqref{eqw:action_full}, the minimal 
coupling to
torsion is equivalent to a coupling of the chiral fermions to the external effective gauge 
field~\eqref{eq:eff_gauge_field}, while for the 
thermodynamics all dependence on the background torsion comes from the 
shift~\eqref{eq:shift} in the equilibrium partition function. 
There is however a crucial physical difference between this and the coupling to a standard external gauge field: the 
torsional gauge field,
and therefore also the torsional chemical potential~$\mu_{S}$, depends 
not only on the torsion tensor but on the metric functions as well
[see eq.~\eqref{eq:muS_spatial_comp}]. This results in additional terms in the constitutive relations of various 
energy-momentum tensor components, that
can be explicitly computed by taking the appropriate functional derivatives of eq.~\eqref{eq:W(1)}. 

Taking this basic fact into account, we compute the torsional corrections to the energy-momentum 
tensor, beginning with the $00$~component
\begin{align}
\langle\Theta_{00}\rangle^{(1)}&\equiv -{T_{0}e^{\sigma}\over \sqrt{g}}{\delta W^{(1)}\over \delta \sigma} 
=-e^{2\sigma}\mu_{S}\left(\langle n\rangle+{d\langle n\rangle\over d\sigma}
\right),
\label{eq:theta00_<n>}
\end{align}
where we have applied that, unlike~$T$
and~$\langle n\rangle$, the torsional chemical potential~$\mu_{S}$ 
is independent of~$\sigma$, as we pointed out after eq.~\eqref{eq:muS_spatial_comp}. 
To recast this expression in terms of the fluid's energy density, we make use the thermodynamic relation
\begin{align}
\varepsilon=-\mathcal{P}+T{\partial\mathcal{P}\over \partial T}
+\mu_{L}{\partial\mathcal{P}\over \partial\mu_{L}}.
\end{align}
Using eq.~\eqref{eq:muLT0defs}, we see that 
the two partial derivatives can be
combined into a single total one with respect to~$\sigma$
\begin{align}
\varepsilon=-\mathcal{P}-{d\mathcal{P}\over d\sigma},
\end{align}
and differentiating this expression with respect to~$\mu_{L}$, we arrive at
\begin{align}
{\partial\varepsilon\over\partial\mu_{L}}&=-\langle n\rangle-{\partial\over\partial\mu_{L}}
\left({d\mathcal{P}\over d\sigma}\right).
\label{eq:partialepsilon/partialmu_previous}
\end{align}
The total and partial derivatives on the
right-hand side, however, do not commute with one
another, but we have instead
\begin{align}
{\partial\over\partial\mu_{L}}
\left({d\mathcal{P}\over d\sigma}\right)
&={d\over d\sigma}\left({\partial\mathcal{P}\over\partial\mu_{L}}\right)
-{\partial\mathcal{P}\over\partial\mu_{L}}
\nonumber \\[0.2cm]
&={d\langle n\rangle\over d\sigma}-\langle n\rangle.
\end{align}
Applying this identity in eq.~\eqref{eq:partialepsilon/partialmu_previous}, we find
\begin{align}
{\partial\varepsilon\over\partial\mu_{L}}=-{d\langle n\rangle\over d\sigma},
\label{eq:depsilondmu=dndsigma}
\end{align}
and arrive at the result
\begin{align}
\langle\Theta_{00}\rangle^{(1)}&
=e^{2\sigma}\mu_{S}\left({\partial\varepsilon\over\partial\mu_{L}}-\langle n\rangle
\right) \nonumber \\[0.2cm]
&
=-{1\over 8}e^{2\sigma}
\epsilon^{ijk}g_{i\ell}\big(T^{\ell}_{\,\,\,jk}-2a_{j}T^{\ell}_{\,\,\,0k}\big)
\left({\partial\varepsilon\over\partial \mu_{L}}-\langle n\rangle\right),
\label{eq:<Theta00>}
\end{align}
after substituting the definition~\eqref{eq:muS_spatial_comp} in the second line.

The mixed components of the energy-momentum tensor, on the other hand, are obtained by taking variations
with respect to the KK gauge field~$a_{i}$ 
\begin{align}
\langle\Theta^{i}_{0}\rangle^{(1)}
&\equiv{T_{0}e^{-\sigma}\over \sqrt{g}}{\delta W^{(1)}\over \delta a_{i}}
={1\over 4}\langle n\rangle\epsilon^{ijk}g_{j\ell}T^{\ell}_{\,\,\,0k}.
\label{eq:Thetaupidown0generalf}
\end{align}
Using the definition~$\langle\Theta^{i}_{0}\rangle=q^{i}u_{0}$, we find the linear 
torsional contribution to the heat current
\begin{align}
q^{i}=-{1\over 4}\langle n\rangle e^{-\sigma}\epsilon^{ijk}g_{j\ell}T^{\ell}_{\,\,\,0k},
\label{eq:heat_current_general_noncov}
\end{align}
with~$q_{0}=0$ (or~$q^{0}=-a_{i}q^{i}$) due to transversality.
Interestingly, this result is obtained from the covariant expression
\begin{align}
q^{\mu}=-{1\over 4}\langle n\rangle \epsilon^{\mu\nu\alpha\beta}G_{\alpha\sigma}
T^{\sigma}_{\,\,\,\lambda\nu}u^{\lambda}u_{\beta},
\label{eq:qmu}
\end{align}
particularized
to the static line element~\eqref{eq:eff_gauge_field} and with the 
equilibrium four-velocity taking the form given in eq.~\eqref{eq:uaomega}. 

We finally compute the spatial components, taking
into account
that~$\sqrt{g}\epsilon^{ijk}$ is
independent of the transverse metric~$g_{ij}$
\begin{align}
\langle\Theta^{ij}\rangle^{(1)}
&\equiv{2T_{0}e^{-\sigma}\over\sqrt{g}}{\delta W^{(1)}\over \delta g_{ij}}
={2\over \sqrt{g}}\langle n\rangle{\partial\over \partial g_{ij}}(\sqrt{g}\mu_{S}).
\label{eq:Thetaij}
\end{align}
The derivative in the last expression 
can be explicitly computed from eq.~\eqref{eq:muS_spatial_comp}, with the result
\begin{align}
\langle\Theta^{ij}\rangle^{(1)}&=
-{1\over 4}\langle n\rangle\epsilon^{k\ell (i}\big(T^{j)}_{\,\,\,\,k\ell}-2a_{k}T^{j)}_{\,\,\,\,0\ell}\big).
\label{eq:Thetaijgeneralf}
\end{align}
The right-hand side is the equilibrium form of the covariant expression
\begin{align}
\mathfrak{T}^{\mu\nu}&={1\over 4}\langle n\rangle 
\epsilon^{\sigma\alpha\beta(\mu}\Delta^{\nu)}_{\lambda}
T^{\lambda}_{\,\,\,\sigma\alpha}u_{\beta},
\end{align} 
written in terms of the projector~$\Delta^{\mu}_{\nu}=\delta^{\mu}_{\nu}+u^{\mu}u_{\nu}$.
The tensor~$\mathfrak{T}^{\mu\nu}$ can actually be split into a trace and a traceless parts 
\begin{align}
\mathfrak{T}^{\mu\nu}=\Pi\Delta^{\mu\nu}+\pi^{\mu\nu},
\end{align}
with
\begin{align}
\Pi&={2\over 3}\langle n\rangle\mu_{S} \nonumber \\[0.2cm]
\pi^{\mu\nu}&={1\over 4}\langle n\rangle \Delta^{\mu\nu}_{\lambda\rho}
\epsilon^{\sigma\alpha\beta\rho}
T^{\lambda}_{\,\,\,\,\sigma\alpha}u_{\beta}.
\label{eq:pimunu_ndiss}
\end{align}
To obtain the first identity we have applied again eq.~\eqref{eq:muS_spatial_comp}, while in the second one
we introduced the symmetric tensor
\begin{align}
\Delta^{\mu\nu}_{\alpha\beta}\equiv\Delta^{\mu}_{(\alpha}\Delta^{\nu}_{\beta)}-{1\over 3}\Delta^{\mu\nu}
\Delta_{\alpha\beta},
\end{align}
which is both transverse and traceless. 

Collecting all previous results together, we find the covariant form of the
linear torsional contributions to the energy-momentum tensor, at first order in the derivative expansion
\begin{align}
\langle \Theta^{\mu\nu}\rangle^{(1)}
=\mu_{S}\left({\partial\varepsilon\over\partial \mu_{L}}-\langle n\rangle\right)u^{\mu}u^{\nu}
+{2\over 3}\mu_{S}\langle n\rangle\Delta^{\mu\nu}
+q^{\mu}u^{\nu}+q^{\nu}u^{\mu}+\pi^{\mu\nu}.
\label{eq:emtensor_full_general}
\end{align}
We notice the existence of nondissipative torsional terms to the heat current, 
dynamical pressure, and anisotropic stress, with their 
corresponding transport coefficients given  
in eqs.~\eqref{eq:qmu} and~\eqref{eq:pimunu_ndiss}. In particular, the term~$\pi^{\mu\nu}$ can be 
interpreted as a four-dimensional counterpart of the
Hall viscosity term in 2+1 dimensions~\cite{Hoyos:2014pba}. Let us recall that in this case
the torsional contribution~$\langle\Theta^{ij}\rangle$
at equilibrium is obtained from the covariant expression~\cite{Valle:2015hfa}
\begin{align}
\mathfrak{T}^{\mu\nu}_{2+1}&=\eta_{H}\sigma_{\beta}^{\,\,\,(\mu}\epsilon^{\nu)\alpha\beta}u_{\alpha},
\label{eq:hall_visco_2+1}
\end{align}
where~$\sigma^{\mu\nu}$ is the shear tensor
\begin{align}
\sigma^{\mu\nu}&=\Delta^{\mu\alpha}\Delta^{\nu\beta}\big(\nabla_{\alpha}u_{\beta}+\nabla_{\beta}u_{\alpha}-\Delta_{\alpha\beta}\nabla_{\sigma}u^{\sigma}\big),
\end{align}
linear in the torsion components. There is however an important difference with respect to the 
four-dimensional result~\eqref{eq:pimunu_ndiss}, since~$\mathfrak{T}^{\mu\nu}_{2+1}$
is traceless and includes no torsional corrections
to the dynamic pressure. 

Another relevant feature of the linear torsional 
corrections~\eqref{eq:emtensor_full_general} is that they
preserve the trace of the energy-momentum tensor. 
Indeed, using the properties of~$q^{\mu}$ and~$\pi^{\mu\nu}$, as 
well as eq.~\eqref{eq:depsilondmu=dndsigma}, we find
\begin{align}
\langle\Theta^{\mu}_{\mu}\rangle^{(1)}
=\mu_{S}\left({d\langle n\rangle\over d\sigma}
+3\langle n\rangle\right),
\end{align}
which vanishes as a consequence of the homogeneity property of the number density
\begin{align}
{d\langle n\rangle\over d\sigma}=-3\langle n\rangle,
\end{align} 
imposed by classical scale invariance. Here we find a further difference with respect to 
the $(2+1)$-dimensional case, where the linear torsional corrections to the trace of the energy-momentum tensor
are given by~\cite{Valle:2015hfa}
\begin{align}
\langle\Theta^{\mu}_{\mu}\rangle^{(1)}_{2+1}&=
e^{2\sigma}\left({d\langle\overline{\Psi}\Psi\rangle\over d\sigma}+2\langle\overline{\Psi}\Psi\rangle
\right)\delta m
-{1\over 2}e^{2\sigma}\langle\overline{\Psi}\Psi\rangle 
u^{\mu}\epsilon^{\nu\alpha\sigma}T_{\alpha\mu\nu}u_{\sigma}.
\label{eq:traceTheta2+1_pre}
\end{align}
This expression does not vanish after imposing the corresponding homogeneity relation for the fermion
condensate
\begin{align}
{d\langle\overline{\Psi}\Psi\rangle\over d\sigma}=-2\langle\overline{\Psi}\Psi\rangle.
\end{align} 
To understand this physically, we recall that 
in $2+1$ dimensions the coupling to
torsion amounts to a shift in the fermion mass~\cite{Valle:2015hfa}, 
which necessarily breaks classical scale invariance.
In the $(3+1)$-dimensional case, by contrast, the effect of 
torsion is equivalent to a coupling to an external
gauge field, which does not conflict with classical scale invariance. 
As a consequence,~$\langle \Theta^{\mu}_{\mu}\rangle_{3+1}$ does 
not receive corrections linear in the torsion.

\paragraph{The spin energy potential.}
Being determined by the variation of the equilibrium partition function
with respect to the torsion tensor~\cite{Hehl:1976kj,Manes:2020zdd}, the
leading order contribution to the spin energy potential comes from the part
of the partition function linear in the torsion and computed
in eq~\eqref{eq:W(1)}
\begin{align}
\langle \Psi_{\mu}^{\,\,\,\,\alpha\beta}\rangle^{(0)}
&\equiv{2T_{0}e^{-\sigma}\over \sqrt{g}}{\delta W^{(1)}\over \delta T^{\mu}_{\,\,\,\alpha\beta}}
=2\langle n\rangle {\partial\mu_{S}\over \partial T^{\mu}_{\,\,\,\,\alpha\beta}},
\label{eq:spin_energ_pot_def}
\end{align}
and is of order zero in the derivative expansion.
Applying eq.~\eqref{eq:muS_spatial_comp}, we compute the nonvanishing components 
of the spin energy potential
\begin{align}
\langle\Psi^{ijk}\rangle^{(0)}
&={1\over 4}\langle n\rangle \epsilon^{ijk}, \nonumber \\[0.2cm]
\langle\Psi^{i0j}\rangle^{(0)}
&=-\langle\Psi^{ij0}\rangle^{(0)}
=-\langle\Psi^{0ij}\rangle^{(0)}
={1\over 4}\langle n\rangle \epsilon^{ijk}a_{k}
=-a_{k}\langle\Psi^{ikj}\rangle^{(0)},
\end{align}
a result that can be obtained from the covariant expression
\begin{align}
\langle\Psi^{\mu\nu\alpha}\rangle^{(0)}
=-{1\over 4}\langle n\rangle\epsilon^{\mu\nu\alpha\beta}u_{\beta},
\label{eq:psi_gen_cov_expression}
\end{align}
evaluated at equilibrium.
The nonvanishing KK-invariant components can be found using the explicit expressions
given in~\cite{Manes:2020zdd}, with the result
\begin{align}
\langle\boldsymbol{\Psi}^{ijk}\rangle^{(0)}
={1\over 4}\langle n\rangle \epsilon^{ijk}.
\label{eq:spin_current_general}
\end{align} 
The linear torsional
contribution to the spin energy potential, on the other hand, come from the second-order correction
to the partition function
\begin{align}
W^{(2)}={1\over 2T_{0}}\int d^{3}x\sqrt{g}\,e^{\sigma}{\partial\langle n\rangle\over\partial\mu_{L}}
\mu_{S}^{2}.
\end{align}
Applying then the definition~\eqref{eq:spin_energ_pot_def}, we find
\begin{align}
\langle\Psi^{\mu}_{\,\,\,\nu\alpha}\rangle^{(1)}
=2\mu_{S}{\partial\langle n\rangle\over\partial\mu_{L}}
{\partial\mu_{S}\over \partial T^{\mu}_{\,\,\,\nu\alpha}}
={1\over 4}\mu_{S}{\partial\langle n\rangle\over\partial\mu_{L}}
\epsilon^{\mu}_{\,\,\,\nu\alpha\beta}u^{\beta}.
\label{eq:spin_energy_potlinT_gen}
\end{align}
This is linear in the torsion and therefore first order in the derivative expansion.

\section{Thermal Weyl fermions in magnetic background torsion}
\label{sec:microscopic}

After having analyzed the general structure of linear torsional terms in the finite temperature
constitutive relations for the
chiral current,
energy-momentum tensor, and spin energy potential, we focus our attention on some 
particular geometrical backgrounds with 
different types of torsion. 
The first to be considered is defined by the tetrad
\begin{align}
e^{a}=\left(dt,dx,dy,dz-{1\over 2}Bydx+{1\over 2}Bxdy\right),
\label{eq:vierbein_torsional}
\end{align}
and the spin connection one-form
\begin{align}
\omega_{12}&=-\omega_{21}=-{1\over 2}(B-\mathcal{T})e^{3}, \nonumber \\[0.2cm]
\omega_{13}&=-\omega_{31}=-{1\over 2}(B-\mathcal{T})e^{2}, 
\label{eq:full_spin_connection_nonzero}\\[0.2cm]
\omega_{23}&=-\omega_{32}={1\over 2}(B-\mathcal{T})e^{1},
\nonumber
\end{align}
depending on two real background data~$B$ and~$\mathcal{T}$. 
A short calculation shows that the latter parametrizes the
background torsion
\begin{align}
T^{a}=\mathcal{T}\delta_{3}^{a}dxdy,
\label{eq:torsion_Stone_etal}
\end{align}
while the curvature takes the form
\begin{align}
R_{12}&=-R_{21}=-{1\over 4}(B-\mathcal{T})(3B-\mathcal{T})e^{1}e^{2},
\nonumber \\[0.2cm]
R_{13}&=-R_{31}={1\over 4}(B-\mathcal{T})^{2}e^{1}e^{3}, \\[0.2cm]
R_{23}&=-R_{32}={1\over 4}(B-\mathcal{T})^{2}e^{2}e^{3}.
\nonumber
\end{align}
This geometry is a generalization of the
one applied in refs.~\cite{Huang:2019haq,Laurila:2020yll} to the study of torsional Landau levels, 
that is recovered when~$B=\mathcal{T}$ and both the spin connection and curvature vanish. 
By including a nonzero
spin connection, however, here we are able to disentangle the ``torsional magnetic field''
parameter~$B$ from the bona fide torsion~$\mathcal{T}$. 

The background~\eqref{eq:vierbein_torsional}-\eqref{eq:full_spin_connection_nonzero} 
describes a fluid in equilibrium with four-velocity~$u\equiv -e^{0}=-dt$ and
vanishing vorticity and acceleration,~$\omega=\mathfrak{a}=0$. 
Implementing the electric-magnetic decomposition of 
the torsion two-form with respect to~$u$ 
\begin{align}
T^{a}\equiv 
\boldsymbol{B}^{a}+uE^{a},
\end{align} 
we see that~\eqref{eq:torsion_Stone_etal} is purely magnetic with~$\boldsymbol{B}^{a}
=\mathcal{T}\delta_{3}^{a}dxdy$,
while the corresponding
decomposition for the vierbein and the spin connection
\begin{align}
e^{a}&=\pmb{\bm{e}}^{a}-\chi^{a}u, \nonumber \\[0.2cm]
\omega^{a}_{\,\,\,b}&=\boldsymbol{\omega}^{a}_{\,\,\,b}-\mu^{a}_{\,\,\,b}u,
\label{eq:spin_connection_EMdecompgen}
\end{align}
shows that~$\chi^{a}=\delta^{a}_{0}$ and
the spin chemical potential is zero,~$\mu^{a}_{\,\,\,b}=0$.
Moreover,
the equilibrium constraint~\cite{Valle:2021nfv}
\begin{align}
E^{a}-\big(\boldsymbol{D}+\mathfrak{a}\big)\chi^{a}+\mu^{a}_{\,\,\,b}\pmb{\bm{e}}^{b}=0,
\label{eq:equilibrium_constr}
\end{align}
with~$\boldsymbol{D}$ the covariant derivative 
defined from the magnetic component
of the spin connection, can be seen to be
trivially satisfied. This shows the independence of the two background data~$B$ and~$\mathcal{T}$.
Finally, the tetrad~\eqref{eq:vierbein_torsional} defines a static line element 
of the form~\eqref{eq:static_metric} 
with~$\sigma=a_{i}=0$
\begin{align}
ds^{2}=-dt^{2}+dx^{2}+dy^{2}+\left[1+{1\over 4}B^{2}(x^{2}+y^{2})\right]dz^{2}
+Bydxdz-Bxdydz.
\label{eq:spatial_metric}
\end{align}
In addition to its minimal coupling to the background geometry just described, we assume  that
the fundamental chiral fermion field is also coupled to an external, purely electrical gauge field of the
form~$\mathcal{A}=-\mu_{L}u$, with~$\mu_{L}$ the chiral chemical potential. 

The fermion Hamiltonian~\eqref{eq:hamiltonian_general} now takes the form
\begin{align}
\mathcal{H}&=i\overline{\sigma}^{k}_{-}\partial_{k}-\mu_{L}-{1\over 4}(\mathcal{T}-B), 
\nonumber\\[0.2cm]
&=\sigma_{x}\left(i\partial_{x}+{i\over 2}By\partial_{z}\right)
+\sigma_{y}\left(i\partial_{y}-{i\over 2}Bx\partial_{z}\right)
+i\sigma_{z}\partial_{z}
-\mu_{L}+{1\over 4}(\mathcal{T}-B).
\label{eq:hamiltonian_stone}
\end{align}
We see that, as in the background analyzed in ref.~\cite{Huang:2019haq},~$B$ parametrizes a sort of external magnetic field coupling to
fermions through their momentum~$p_{z}$, whereas
the torsion~$\mathcal{T}$ enters the Hamiltonian only through a shift in the chiral chemical
potential. For later convenience, we write
\begin{align}
\mathcal{H}\equiv H-\widetilde{\mu},
\label{eq:H_oper}
\end{align}
where the shifted chemical potential~$\widetilde{\mu}$ is defined by
\begin{align}
\widetilde{\mu}\equiv \mu_{L}-{1\over 4}(\mathcal{T}-B).
\label{eq:mutilde_def}
\end{align}

\paragraph{The thermal one-particle Green function.}
To compute the constitutive relations at finite temperature,
we use the imaginary time formalism where the Euclidean time is
compactified to a circle of length~$\beta_{0}=T_{0}^{-1}$ and fermions 
are antiperiodic along the thermal circle. 
More specifically, we work with the fermion Green function
\begin{align}
\mathcal{G}_{ab}(\tau-\tau';\boldsymbol{r},\boldsymbol{r}')\equiv -\langle 
T_{\tau}[\psi_{a}(\tau,\boldsymbol{r})
\psi_{b}(\tau',\boldsymbol{r}')^{\dagger}]\rangle_{\beta_{0}},
\end{align}
where~$T_{\tau}[\ldots]$ 
represents (Euclidean) time ordering. 

Since we are interested in computing
thermal averages in a system of microscopic chiral fermions interacting only with external sources, 
it suffices to consider the one-particle Green function, defined by the matrix equation
\begin{align}
\big(\partial_{\tau}+\mathcal{H}\big)\boldsymbol{\mathcal{G}}(\tau-\tau';\boldsymbol{r},\boldsymbol{r}')=
{1\over \sqrt{-G}}(\overline{\sigma}_{-}^{0})^{-1}
\delta(\tau'-\tau)\delta^{(3)}(\boldsymbol{r}'-\boldsymbol{r}),
\label{eq:green_function_def}
\end{align}
where the Hamiltonian~$\mathcal{H}$ is acting on the $\boldsymbol{r}$~coordinate. 
It is convenient to recast the Green function in terms of 
its Euclidean time Fourier transform
\begin{align}
\mathcal{G}_{ab}(\tau-\tau';\boldsymbol{r},\boldsymbol{r}')
={1\over\beta_{0}}\sum_{n\in\mathbb{Z}}e^{-i\omega_{n}(\tau-\tau')}
\mathcal{G}_{ab}(\omega_{n};\boldsymbol{r},\boldsymbol{r}'),
\label{eq:green_function_time}
\end{align}
with~$\omega_{n}$ the Matsubara frequencies
\begin{align}
\omega_{n}={2\pi\over \beta_{0}}\left(n+{1\over 2}\right).
\label{eq:matsubara_freq}
\end{align}
The new matrix
function~$\boldsymbol{\mathcal{G}}(\omega_{n};\boldsymbol{r},\boldsymbol{r}')$ then satisfies the equation
\begin{align}
\big(i\omega_{n}-\mathcal{H}\big)\boldsymbol{\mathcal{G}}(\omega_{n};\boldsymbol{r},\boldsymbol{r}')
=-{1\over \sqrt{-G}}(\overline{\sigma}_{-}^{0})^{-1}\delta^{(3)}(\boldsymbol{r}-\boldsymbol{r}'),
\label{eq:GF_def_eom_frequencies1}
\end{align}
whose solution gives the
thermal averages of fermion bilinears via the identity~\cite{AGD,Fetter_Walecka,Bruus_Frensberg}
\begin{align}
\langle \psi^{\dagger}\boldsymbol{A}\psi\rangle
={1\over \beta_{0}}\sum_{n\in\mathbb{Z}}{\rm tr\,}\big[\boldsymbol{A}\boldsymbol{\mathcal{G}}(\boldsymbol{r},\boldsymbol{r};\omega_{n})\big],
\label{eq:thermal_average_GF}
\end{align}
with~$\boldsymbol{A}$ an arbitrary linear operator.

In fact,
the thermal one-particle Green function can be expressed in terms of 
the heat kernel associated to the square of the operator~$H$ defined in~\eqref{eq:H_oper}
\begin{align}
\left({\partial\over\partial s}+H^{2}\right)\boldsymbol{\mathcal{K}}(\boldsymbol{r},\boldsymbol{r}';s)=0,
\label{eq:heat_kernel_eq}
\end{align}
where~$\boldsymbol{\mathcal{K}}(\boldsymbol{r},\boldsymbol{r}';s)$ is a $2\times 2$ matrix satisfying the 
boundary condition
\begin{align}
\boldsymbol{\mathcal{K}}(\boldsymbol{r},\boldsymbol{r}';0)
=-{1\over \sqrt{-G}}(\overline{\sigma}_{-}^{0})^{-1}\delta^{(3)}(\boldsymbol{r}-\boldsymbol{r}').
\label{eq:bdy_cond_kcal}
\end{align} 
Using~\eqref{eq:heat_kernel_eq}, it is straightforward to check that the function
\begin{align}
\boldsymbol{\mathcal{G}}(\boldsymbol{r},\boldsymbol{r'};\omega_{n})\equiv 
\int_{0}^{\infty}ds\,e^{-s(\omega_{n}-i\widetilde{\mu})^{2}}\big(-i\omega_{n}-\widetilde{\mu}
-H\big)\boldsymbol{\mathcal{K}}(\boldsymbol{r},\boldsymbol{r'};s),
\label{eq:green_function_nonzero_mu}
\end{align}
solves indeed eq.~\eqref{eq:GF_def_eom_frequencies1}.

Being interested in corrections linear in the torsion, we would only need to solve 
eq.~\eqref{eq:GF_def_eom_frequencies1}
in perturbation theory at first order in~$\mathcal{T}$, as we will actually do in forthcoming sections. 
In the case of the background~\eqref{eq:vierbein_torsional}-\eqref{eq:torsion_Stone_etal}, however,
both the heat kernel~$\boldsymbol{\mathcal{K}}(\boldsymbol{r},\boldsymbol{r}';s)$ and the
one-particle Green function~$\boldsymbol{\mathcal{G}}(\boldsymbol{r},\boldsymbol{r'};\omega_{n})$ 
can be computed exactly.
Using the explicit expression of~$H$ extracted from eq.~\eqref{eq:hamiltonian_stone}, 
we find the following solution to
the heat kernel equation~\eqref{eq:heat_kernel_eq}
(see also ref.~\cite{sondheimer_wilson}) 
\begin{align}
\mathcal{K}_{11}(\boldsymbol{r},\boldsymbol{r}';s)&=\int_{-\infty}^{\infty}{dk_{z}\over 8\pi^{2}}
{Bk_{z}\over \sinh{(Bk_{z}s)}}
\exp\left\{-{iBk_{z}\over 2}\big(xy'-yx'\big)\right. \nonumber \\[0.2cm]
&-\left.{Bk_{z}\over 4}\coth(Bk_{z}s)
\Big[(x-x')^{2}+(y-y')^{2}\Big]-s\big(k_{z}^{2}-Bk_{z}\big)+ik_{z}(z-z')\right\}, \nonumber \\[0.2cm]
\mathcal{K}_{22}(\boldsymbol{r},\boldsymbol{r}';s)&=\int_{-\infty}^{\infty}{dk_{z}\over 8\pi^{2}}
{Bk_{z}\over \sinh{(Bk_{z}s)}}
\exp\left\{-{iBk_{z}\over 2}\big(xy'-yx'\big)\right. 
\label{eq:heat_kernel_final} \\[0.2cm]
&-\left.{Bk_{z}\over 4}\coth(Bk_{z}s)
\Big[(x-x')^{2}+(y-y')^{2}\Big]-s\big(k_{z}^{2}+Bk_{z}\big)+ik_{z}(z-z')\right\},
\nonumber
\end{align}
with~$\mathcal{K}_{12}(\boldsymbol{r},\boldsymbol{r}';s)=\mathcal{K}_{21}(\boldsymbol{r},\boldsymbol{r}';s)=0$. Substituting
these expressions
into eq.~\eqref{eq:green_function_nonzero_mu}, the explicit form of the
Green-function to be used in the computation of thermal 
torsional corrections of the various currents is obtained.

Let us proceed with the evaluation of the thermal average of the chiral 
current~$J^{\mu}\equiv -\psi^{\dagger}\sigma^{\mu}_{-}\psi$. 
Since it offers a template for future analyses, here 
we present the calculations
in some detail. 
We start with the time component
\begin{align}
\langle J^{0}\rangle=-\langle\psi^{\dagger}\sigma^{0}_{-}\psi\rangle=\langle\psi^{\dagger}\psi\rangle,
\end{align}
that at zeroth order in the background data
gives the number density~$\langle n\rangle$, 
the main ingredient
of the general expressions computed in section~\ref{sec:general_analysis}. 
Applying eq.~\eqref{eq:thermal_average_GF} and
the expression of the one-particle Green function in terms of the heat 
kernel~\eqref{eq:green_function_nonzero_mu}, we find after some algebra 
\begin{align}
\langle J^{0}\rangle
&={1\over \beta_{0}}
\sum_{n\in\mathbb{Z}}{\rm tr\,}\boldsymbol{\mathcal{G}}(\boldsymbol{r},\boldsymbol{r};\omega_{n})
\nonumber \\[0.2cm]
&={i\beta_{0}\over 32\pi^{7\over 2}}
\int_{0}^{\infty}ds\,s^{-{3\over 2}}
\theta_{4}'\left(\left.{\widetilde{\mu}\beta_{0}\over 2\pi i}\right|{i\beta_{0}^{2}\over 4\pi s}\right)
\int_{-\infty}^{\infty}dk_{z}\,Bk_{z}\coth{(Bk_{z}s)}
e^{-sk_{z}^{2}} \label{eq:<psidaggerpsi>_second} \\[0.2cm]
&-{B\over 16\pi^2}\int_{\Lambda^{-2}}^{\infty}{ds \over s^{2}}\,
\theta_{4}\left(\left.{\widetilde{\mu}\beta_{0}\over 2\pi i}\right|{i\beta_{0}^{2}\over 4\pi s}\right),
\nonumber
\end{align}
where the sums over Matsubara frequencies have been packed into Jacobi theta functions~\cite{ErdelyiVol2}
\begin{align}
\sum_{n\in\mathbb{Z}}e^{-s(\omega_{n}-i\widetilde{\mu})^{2}}
&={\beta_{0}\over \sqrt{4\pi s}}\theta_{4}\left(\left.
{\widetilde{\mu}\beta_{0}\over 2\pi i}\right|
{i\beta_{0}^{2}\over 4\pi s}\right), \nonumber \\[0.2cm]
\sum_{n\in\mathbb{Z}}(\omega_{n}-i\widetilde{\mu})e^{-s(\omega_{n}-i\widetilde{\mu})^{2}}
&=-{\beta_{0}^{2}\over 8\pi^{3\over 2}} s^{-{3\over 2}}\theta'_{4}\left(\left.
{\widetilde{\mu}\beta_{0}\over 2\pi i}\right|
{i\beta_{0}^{2}\over 4\pi s}\right),
\end{align}
and the prime, as usual, indicates differentiation with respect to the first argument.
In addition, the last integral in eq.~\eqref{eq:<psidaggerpsi>_second} 
has been regularized with a proper time UV cutoff, while
the one in the second line 
is finite due to the absence of a zero mode in~$\theta_{4}'(z|\tau)$.
At this point, we carry out the expansion in~$B$ and~$\mathcal{T}$, keeping only terms linear
in both quantities
\begin{align}
\langle J^{0}\rangle
&={i\beta_{0}\over 32\pi^{3}}
\int_{0}^{\infty}{ds\over s^{3}}
\theta_{4}'\left(\left.{\mu_{L}\beta_{0}\over 2\pi i}\right|{i\beta_{0}^{2}\over 4\pi s}\right)
+{\beta_{0}^{2}\over 256\pi^{4}}\big(B-\mathcal{T}\big)
\int_{0}^{\infty}{ds\over s^{3}}
\theta_{4}''\left(\left.{\mu_{L}\beta_{0}\over 2\pi i}\right|{i\beta_{0}^{2}\over 4\pi s}\right)
\nonumber \\[0.2cm]
&+{B\over 16\pi^2}\int_{\Lambda^{-2}}^{\infty}{ds \over s^{2}}\,
\theta_{4}\left(\left.{\mu_{L}\beta_{0}\over 2\pi i}\right|{i\beta_{0}^{2}\over 4\pi s}\right).
\label{eq:<psi+psi>_second}
\end{align}
Applying then the identity (see Appendix~\ref{app:magic_identity})
\begin{align}
\int_{\Lambda^{-2}}^{\infty}{ds\over s^{n}}\theta_{4}&\left(\left.
{\mu\beta_{0}\over 2\pi i}\right|
{i\beta_{0}^{2}\over 4\pi s}\right) \nonumber \\[0.2cm]
&+{\beta_{0}^{2}\over 16(n-1)\pi^{2}}\int_{0}^{\infty}{dS\over s^{n+1}}\theta_{4}''\left(\left.
{\mu\beta_{0}\over 2\pi i}\right|
{i\beta_{0}^{2}\over 4\pi s}\right)
=
{1\over n-1}\Lambda^{2n-2}
+\mathcal{O}(e^{-\beta_{0}^{2}\Lambda^{2}}),
\label{eq:magic_identity}
\end{align}
valid for~$n>1$, 
we find that all dependence on~$B$ cancels from the cutoff-independent part
\begin{align}
\langle J^{0}\rangle
&={B\over 16\pi^2}\Lambda^{2}
+{i\beta_{0}\over 32\pi^{3}}
\int_{0}^{\infty}{ds\over s^{3}}
\theta_{4}'\left(\left.{\mu_{L}\beta_{0}\over 2\pi i}
\right|{i\beta_{0}^{2}\over 4\pi s}\right) \nonumber \\[0.2cm]
&-{\beta_{0}^{2}\mathcal{T}\over 256\pi^{4}}
\int_{0}^{\infty}{ds\over s^{3}}
\theta_{4}''\left(\left.{\mu_{L}\beta_{0}\over 2\pi i}\right|{i\beta_{0}^{2}\over 4\pi s}\right).
\end{align}
The remaining integrals are computed using the expansion
\begin{align}
\theta_{4}(z|\tau)=1+2\sum_{n=0}^{\infty}(-1)^{n}e^{i\pi n^{2}\tau}\cos(2\pi nz),
\label{eq:theta4expansion}
\end{align} 
and expressed in terms of polylogarithms give
\begin{align}
\langle J^{0}\rangle
&={\Lambda^{2}B\over 16\pi^2}
-{1\over \pi\beta_{0}^{3}}
\big[{\rm Li}_{3}(-e^{-\mu_{L}\beta_{0}})
-{\rm Li}_{3}(-e^{\mu_{L}\beta_{0}})\big] \nonumber \\[0.2cm]
&+{\mathcal{T}\over 8\pi^{2}\beta_{0}^{2}}\big[{\rm Li}_{2}(-e^{-\mu_{L}\beta_{0}})
+{\rm Li}_{2}(-e^{\mu_{L}\beta_{0}})\big].
\end{align}
Finally, using the identity~\cite{Lewin}
\begin{align}
{\rm Li}_{n}(z)+(-1)^{n}{\rm Li}_{n}(z^{-1})=-{(2\pi i)^{n}\over n!}B_{n}\left({\log{z}\over 2\pi i}\right),
\label{eq:poly_identity}
\end{align}
with~$B_{n}(x)$ the $n$-th Bernoulli polynomial, 
we arrive at the result
\begin{align}
\langle J^{0}\rangle
&={\Lambda^{2}B\over 16\pi^2}+{1\over 6}T_{0}^{2}\mu_{L}+{1\over 6\pi^{2}}\mu_{L}^{3}
-\mathcal{T}\left(
{1\over 24}T_{0}^{2}+{1\over 8\pi^{2}}\mu_{L}^{2}\right).
\label{eq:munonzero_trGfinal}
\end{align}
As already pointed out, the only dependence on the parameter~$B$ 
appears on the vacuum contribution.
The terms independent of either~$B$ or~$\mathcal{T}$ give the number density of a chiral
fermion at finite temperature with nonzero chiral chemical potential~\cite{Jaiswal:2015mxa}
\begin{align}
\langle n\rangle\equiv \langle J^{0}\rangle^{(0)}
={1\over 6}\left(T_{0}^{2}\mu_{L}+{1\over \pi^{2}}\mu_{L}^{3}\right),
\label{eq:micro_<n>_explicit}
\end{align} 
which is the quantity to be used as one of the main ingredients of 
the general expressions obtained in section~\ref{sec:general_analysis}.

The computation of the current spatial components
\begin{align}
\langle J^{i}\rangle\equiv\langle\psi^{\dagger}\sigma_{-}^{i}\psi\rangle=
{1\over \beta_{0}}\sum_{n\in\mathbb{Z}}{\rm tr\,}
\big[\sigma_{-}^{i}\boldsymbol{\mathcal{G}}(\boldsymbol{r},\boldsymbol{r};\omega_{n})\big],
\end{align}
is carried out along
similar lines. An evaluation of the relevant traces
\begin{align}
{\rm tr\,}
\big[\sigma_{x}\boldsymbol{\mathcal{G}}(\boldsymbol{r},\boldsymbol{r};\omega_{n})\big]=
{\rm tr\,}
\big[\sigma_{y}\boldsymbol{\mathcal{G}}(\boldsymbol{r},\boldsymbol{r};\omega_{n})\big]=
{\rm tr\,}
\big[\sigma_{z}\boldsymbol{\mathcal{G}}(\boldsymbol{r},\boldsymbol{r};\omega_{n})\big]=
0,
\end{align} 
leads to the conclusion that all three components are equal to zero
\begin{align}
\langle J^{i}\rangle&=0. 
\label{eq:result_chiral_current_stoneetal}
\end{align}
This result, together with the torsional terms 
in eq.~\eqref{eq:munonzero_trGfinal}, show total agreement
with the general expression for the torsional corrections to the current components
given in~\eqref{eq:J_0_general} and~\eqref{eq:J^i_general}, taking into account that
in our background
\begin{align}
\mu_{S}=-{1\over 4}\mathcal{T},
\label{eq:muS_Stone}
\end{align}
as it can be checked from eqs.~\eqref{eq:muS_spatial_comp} and~\eqref{eq:torsion_Stone_etal}.

Next, we apply the thermal one-particle Green function computed above to
evaluate the thermal linear torsional corrections to the
energy-momentum tensor
\begin{align}
\Theta^{\mu\nu}
={i\over 4}\Big\{
\psi^{\dagger}\Big[\overline{\sigma}_{-}^{\mu}
\big(\overrightarrow{\nabla}^{\nu}-i\mathcal{A}^{\nu}\big)
-\big(\overleftarrow{\nabla}^{\nu}+i\mathcal{A}^{\nu}\big)
\overline{\sigma}_{-}^{\mu}+(\mu\leftrightarrow\nu)\Big]\psi
\Big\},
\label{eq:emtensor_full_symmetric}
\end{align}
beginning with
\begin{align}
\langle \Theta_{00}\rangle
&=-{i\over \beta_{0}}
\sum_{n\in\mathbb{Z}}{\rm tr\,}\big[(\omega_{n}-i\mu_{L})
\boldsymbol{\mathcal{G}}(\boldsymbol{r},\boldsymbol{r};
\omega_{n})\big],
\end{align}
where in writing this expression we took into account 
that~$\omega^{a}_{\,\,\,\,b0}=0$~[cf.~\eqref{eq:full_spin_connection_nonzero}].
The calculation is similar to the one of the chiral current,
leading also to a cancellation of all terms proportional to~$B$
\begin{align}
\langle \Theta_{00}\rangle
={\Lambda^{4}\over 32\pi^{2}}+{1\over 120}\left(7\pi^{2}T_{0}^{4}
+30T_{0}^{2}\mu_{L}^{2}+{15\over \pi^{2}}\mu_{L}^{4}\right)
-{\mathcal{T}\over 12}\left(T_{0}^{2}\mu_{L}+{1\over \pi^{2}}\mu_{L}^{3}\right).
\label{eq:Theta00_stone}
\end{align} 
From the finite,~$\mathcal{T}$-independent terms we read the perfect fluid energy 
density~\cite{Jaiswal:2015mxa}
\begin{align}
\varepsilon={1\over 120}\left(7\pi^{2}T_{0}^{4}
+30T_{0}^{2}\mu_{L}^{2}+{15\over \pi^{2}}\mu_{L}^{4}\right).
\label{eq:energy_density_sb}
\end{align}
Using~\eqref{eq:muS_Stone}, as well as the expression of~$\langle n\rangle$ obtained in 
eq.~\eqref{eq:micro_<n>_explicit}, we find that the
linear torsional terms in~\eqref{eq:Theta00_stone} reproduce the result 
derived from eq.~\eqref{eq:<Theta00>}.

The remaining diagonal components
\begin{align}
\langle\Theta_{ii}\rangle&={i\over 2\beta_{0}}\sum_{n\in \mathbb{Z}}{\rm tr\,}\big[
\overline{\sigma}_{-i}\big(\nabla_{i}-\nabla'_{i}\big)\boldsymbol{\mathcal{G}}(\omega_{n},\boldsymbol{r},\boldsymbol{r}')
\big]\Big|_{\boldsymbol{r}'\rightarrow\boldsymbol{r}},
\end{align}
are computed similarly, now including
the nonzero contributions from the spin connection~\eqref{eq:full_spin_connection_nonzero} 
in the covariant derivatives. These
terms, proportional to~$\mathcal{T}-B$, cancel those
scaling with~$B$ in the one-particle thermal Green function, rendering 
again $B$-independent expressions
\begin{align}
\langle \Theta_{11}\rangle&=\langle \Theta_{22}\rangle
=-{\Lambda^{4}\over 32\pi^{2}}+{1\over 360}\left(7\pi^{2}T_{0}^{4}
+30 T_{0}^{2}\mu_{L}^{2}+{15\over \pi^{2}}\mu_{L}^{4}\right), \nonumber \\[0.2cm]
\langle \Theta_{33}\rangle
&=-{\Lambda^{4}\over 32\pi^{2}}+{1\over 360}\left(7\pi^{2}T_{0}^{4}
+30 T_{0}^{2}\mu_{L}^{2}+{15\over \pi^{2}}\mu_{L}^{4}\right)
-{\mathcal{T}\over 12}\left(T_{0}^{2}\mu_{L}+{1\over \pi^{2}}\mu_{L}^{3}\right).
\label{eq:Theta_22_33_stone}
\end{align}
We see that only~$\langle\Theta_{33}\rangle$ picks up corrections linear in the torsion, which 
are proportional
to~$\langle n\rangle$ as given in eq.~\eqref{eq:micro_<n>_explicit}. This 
is totally consistent with~\eqref{eq:Thetaijgeneralf},
taking into account that in our background~$T^{i}_{\,\,jk}=0$ for~$i=1,2$ or~$j,k=3$.
Moreover, from the finite $\mathcal{T}$-independent terms we read the perfect fluid 
pressure~\cite{Jaiswal:2015mxa}
\begin{align}
\mathcal{P}&={1\over 360}\left(7\pi^{2}T_{0}^{4}
+30 T_{0}^{2}\mu_{L}^{2}+{15\over \pi^{2}}\mu_{L}^{4}\right).
\label{eq:Theta11_stone}
\end{align}
Notice that, unlike the case of the chiral current, the vacuum contributions to the 
diagonal components of the energy-momentum tensor do
not depend on the background data.

As for the off-diagonal terms, omitting the vacuum part, we find that the only nonzero components are
\begin{align}
\langle \Theta_{13}\rangle&=\langle \Theta_{31}\rangle
=-{By\over 2}\left[
{1\over 360}\left(7\pi^{2}T_{0}^{4}
+30\mu_{L}^{2}T_{0}^{2}+{15\over \pi^{2}}\mu_{L}^{4}\right)\right], 
\nonumber \\[0.2cm]
\langle \Theta_{23}\rangle&=\langle \Theta_{32}\rangle=
{Bx\over 2}\left[
{1\over 360}\left(7\pi^{2}T_{0}^{4}
+30\mu_{L}^{2}T_{0}^{2}+{15\over \pi^{2}}\mu_{L}^{4}\right)\right],
\label{eq:off-diagonal_stoneimpr}
\end{align}
all independent of the torsion parameter~$\mathcal{T}$. In fact, the
components just computed are part of the perfect-fluid energy-momentum tensor
\begin{align}
\langle \Theta^{\rm (pf)}_{\mu\nu}\rangle=\big(\varepsilon+\mathcal{P}\big)u_{\mu}u_{\nu}
+\mathcal{P}G_{\mu\nu}.
\label{eq:pf_em_tensor_gen}
\end{align}
Indeed, using 
eq.~\eqref{eq:spatial_metric} we find
\begin{align}
\langle \Theta_{13}^{\rm (pf)}\rangle&=\langle \Theta_{31}^{\rm (pf)}\rangle
=\mathcal{P}G_{13}, \nonumber \\[0.2cm]
\langle \Theta_{23}^{\rm (pf)}\rangle&=\langle \Theta_{32}^{\rm (pf)}\rangle
=\mathcal{P}G_{23},
\end{align}
which, together with~\eqref{eq:Theta11_stone}, reproduces~\eqref{eq:off-diagonal_stoneimpr}.
In addition, 
since the~$B$-dependent terms in~$G_{33}$ are of second order in this 
parameter,~$\mathcal{P}G_{33}$ does not contribute any linear correction 
to~$\langle \Theta_{33}\rangle$ computed in
the second identity of eq.~\eqref{eq:Theta_22_33_stone}. To compare with 
the expressions derived in sec.~\ref{sec:general_analysis}, we should keep in mind 
that~$T^{i}_{\,\,0j}=0$ in the background of interest, so
the absence of torsional corrections found in the microscopic calculation  
of the off-diagonal components matches with the result of applying eqs.~\eqref{eq:Thetaijgeneralf} and~\eqref{eq:Thetaupidown0generalf}.

To close this section, 
let us compute the thermal-averaged spin energy potential. 
Taking the variation of the fermionic action~\eqref{eq:action_effective_axial} with
respect to the torsion tensor~$T^{\mu}_{\,\,\,\alpha\beta}$, 
we find that the spin energy potential is expressed 
in term of the chiral current as
\begin{align}
\Psi^{\mu\nu\alpha}={1\over 4}\epsilon^{\mu\nu\alpha\beta}J_{\beta}.
\end{align} 
Inserting eqs.~\eqref{eq:micro_<n>_explicit} and~\eqref{eq:result_chiral_current_stoneetal}, 
the only nonzero components are found to be
\begin{align}
\langle\Psi^{ijk}\rangle^{(0)}
&={1\over 24}\left(T_{0}^{2}\mu_{L}+{1\over \pi^{2}}\mu_{L}^{3}\right)\epsilon^{ijk},
\end{align}
in accord with the result shown in eq.~\eqref{eq:spin_current_general}. 
The contribution
linear in~$\mathcal{T}$ is obtained from the corresponding terms in~\eqref{eq:munonzero_trGfinal}
\begin{align}
\langle\Psi^{ijk}\rangle^{(1)}
&=-{1\over 96}\mathcal{T}\left(T_{0}^{2}+{3\over \pi^{2}}\mu_{L}^{2}\right)\epsilon^{ijk},
\end{align} 
and also agrees with eq.~\eqref{eq:spin_energy_potlinT_gen}
evaluated in the present background\footnote{Notice that since~$a_{i}=0$ there is no difference
between the components~$\langle\Psi^{ijk}\rangle$ and 
their KK-invariant counterparts~$\langle\boldsymbol{\Psi}^{ijk}
\rangle$.}. 

\section{Disentangling torsion from vorticity}
\label{sec:T0ij}

The second torsional background to be studied also has magnetic torsion
and is closely related to the geometry
studied in ref.~\cite{Khaidukov:2018oat}.
Its vierbein takes the form
\begin{align}
e^{a}=\left(dt-{1\over 2}Bydx+{1\over 2}Bxdy,dx,dy,dz\right),
\label{eq:vckzgen}
\end{align}
with~$B$ a real constant, while the spin connection has the following nonzero components
\begin{align}
\omega_{01}&=-\omega_{10}={1\over 2}(\mathcal{T}_{B}-B)e^{2}, \nonumber \\[0.2cm]
\omega_{02}&=-\omega_{20}=-{1\over 2}(\mathcal{T}_{B}-B)e^{1}, \label{eq:sckzgen}\\[0.2cm]
\omega_{12}&=-\omega_{21}=-{1\over 2}(\mathcal{T}_{B}-B)e^{0}. \nonumber 
\end{align}
The curvature is computed to be nonzero
\begin{align}
R_{01}&=-R_{10}=-{1\over 4}(\mathcal{T}_{B}-B)^{2}e^{0}e^{1}, \nonumber \\[0.2cm]
R_{02}&=-R_{20}=-{1\over 4}(\mathcal{T}_{B}-B)^{2}e^{0}e^{2}, \\[0.2cm]
R_{12}&=-R_{21}={1\over 2}B(\mathcal{T}_{B}-B)e^{1}e^{2},
\nonumber
\end{align}
and of second order in~$B$ and~$\mathcal{T}_{B}$, while the purely magnetic torsion
\begin{align}
T^{a}=\mathcal{T}_{B}\delta^{a}_{0}dxdy,
\label{eq:kzgen_magtorsion}
\end{align}
is linear in~$\mathcal{T}_{B}$. 
This geometry describes a fermion fluid with velocity
\begin{align}
u=-e^{0}=-dt+{1\over 2}Bydx-{1\over 2}Bxdy,
\label{eq:velocity_oneform_BTB}
\end{align} 
and the following values for the fluid's vorticity and acceleration
\begin{align}
\omega&=-{1\over 2}Bdxdy, \nonumber \\[0.2cm]
\mathfrak{a}&=0.
\end{align}
These expressions provide the interpretation of the 
two background data~$B$ and~$\mathcal{T}_{B}$ as respectively parametrizing vorticity and torsion.  
Carrying out the electric-magnetic decompositions shown in eq.~\eqref{eq:spin_connection_EMdecompgen},
we find that~$\chi^{a}=\delta^{a}_{0}$ and the fluid has a nonzero spin chemical
potential,~$\mu_{12}=-\mu_{21}=-{1\over 2}(\mathcal{T}_{B}-B)$.
The equilibrium constraint~\eqref{eq:equilibrium_constr} 
is also identically satisfied,
implying also here the independence of~$B$ and~$\mathcal{T}_{B}$.

Unlike in the case of
the geometry studied in the previous section, now the Green function equation cannot be solved exactly, so
we rely on perturbation theory to compute thermal corrections linear in~$B$ and~$\mathcal{T}_{B}$.
We thus split the Hamiltonian~\eqref{eq:hamiltonian_general} particularized to this background as
\begin{align}
\mathcal{H}
=\mathcal{H}^{(0)}+\mathcal{H}^{(1)},
\end{align}
where~$\mathcal{H}^{(0)}$ is its free part 
\begin{align}
\mathcal{H}^{(0)}&=i\sigma_{x}\partial_{x}+i\sigma_{y}\partial_{y}+i\sigma_{z}\partial_{z}
-\mu_{L}\mathbb{1},
\label{eq:freehamiltonian_fermion}
\end{align}
and
\begin{align}
\mathcal{H}^{(1)}&={i\over 2}B\mathbb{1}\big(y\partial_{x}-x\partial_{y}\big)
+{1\over 2}Bx\sigma_{x}\partial_{z}+{1\over 2}By\sigma_{y}\partial_{z} \nonumber \\[0.2cm]
&+\sigma_{z}\left[{1\over 4}(\mathcal{T}_{B}-B)-{1\over 2}B\big(x\partial_{x}+y\partial_{y}\big)\right],
\label{eq:kzgeninthamiltonian}
\end{align}
describes the interaction of the
microscopic Weyl fermions with the background geometry linear in the background data.
Equation~\eqref{eq:GF_def_eom_frequencies1} then takes the form 
\begin{align}
\big(i\omega_{n}\mathbb{1}-\mathcal{H}^{(0)}-\mathcal{H}^{(1)}\big)
\boldsymbol{\mathcal{G}}(\omega_{n};\boldsymbol{r},\boldsymbol{r}')
=\left(\mathbb{1}-{1\over 2}By\sigma_{x}+{1\over 2}Bx\sigma_{y}\right)
\delta^{(3)}(\boldsymbol{r}-\boldsymbol{r}'),
\label{eq:kzgenGF0}
\end{align}
and is solved by splitting the Green function into a free part and a perturbation
\begin{align}
\boldsymbol{\mathcal{G}}(\omega_{n};\boldsymbol{r},\boldsymbol{r}')
=\boldsymbol{\mathcal{G}}^{(0)}(\omega_{n};\boldsymbol{r},\boldsymbol{r}')
+\boldsymbol{\mathcal{G}}^{(1)}(\omega_{n};\boldsymbol{r},\boldsymbol{r}'),
\label{eq:first_eq_pert_th}
\end{align}
where~$\boldsymbol{\mathcal{G}}^{(0)}(\omega_{n};\boldsymbol{r},\boldsymbol{r}')$ 
satisfies the equation
\begin{align}
\big(i\omega_{n}\mathbb{1}-\mathcal{H}^{(0)}\big)\boldsymbol{\mathcal{G}}^{(0)}(\omega_{n};\boldsymbol{r},
\boldsymbol{r}')=\delta^{(3)}(\boldsymbol{r}-\boldsymbol{r}'),
\end{align}
whose solution is given by
\begin{align}
\boldsymbol{\mathcal{G}}^{(0)}(\omega_{n};\boldsymbol{r},&\boldsymbol{r'})  
\label{eq:GF_zero_order}\\[0.2cm]
&=-{i\over 16\pi^{3/2}}\int_{0}^{\infty}ds\,s^{-{5\over 2}}e^{-s(\omega_{n}-i\mu)^{2}}
e^{-{1\over 4s}(\boldsymbol{r}-\boldsymbol{r}')^{2}}
\big[2s\big(\omega_{n}-i\mu_{L}\big)\mathbb{1}+\big(\boldsymbol{r}-\boldsymbol{r}'\big)
\cdot\boldsymbol{\sigma}\big].
\nonumber
\end{align}
The linear correction~$\boldsymbol{\mathcal{G}}^{(1)}(\omega_{n};\boldsymbol{r},\boldsymbol{r}')$, on the other hand,
takes the form
\begin{align}
\boldsymbol{\mathcal{G}}^{(1)}(\omega_{n};\boldsymbol{r},\boldsymbol{r}')
&=\int d^{3}\boldsymbol{r}''\,\boldsymbol{\mathcal{G}}^{(0)}(\omega_{n};\boldsymbol{r},\boldsymbol{r}'')
\mathcal{H}^{(1)}\boldsymbol{\mathcal{G}}^{(0)}(\omega_{n};\boldsymbol{r}'',\boldsymbol{r}')
\nonumber \\[0.2cm]
&-{1\over 2}B\boldsymbol{\mathcal{G}}^{(0)}(\omega_{n};\boldsymbol{r},\boldsymbol{r'})
\big(x'\sigma_{y}-y'\sigma_{x}\big),
\label{eq:last_eq_pert_th}
\end{align}
where the perturbation Hamiltonian~$\mathcal{H}^{(1)}$ 
acts on the~$\boldsymbol{r}''$ argument of the free Green function to its right, and the 
last term originates in the piece linear in~$B$ of 
the right-hand side of eq.~\eqref{eq:kzgenGF0}. 

The terms linear in the background data in the finite temperature constitutive relation of the
chiral current are given by
\begin{align}
\langle J^{\mu}\rangle^{(1)}=-{1\over\beta_{0}}\sum_{n\in\mathbb{Z}}{\rm tr\,}\Big[\sigma^{\mu}_{-}
\boldsymbol{\mathcal{G}}^{(1)}(\omega_{n};\boldsymbol{r},\boldsymbol{r})\Big].
\end{align}
Applying the same techniques as in the previous section, we find
\begin{align}
\langle J_{0}\rangle^{(1)}=\langle J^{1}\rangle^{(1)}=\langle J^{2}\rangle^{(2)}=0,
\label{eq:kzgen_J0(1)final}
\end{align}
where the vanishing of~$\langle J^{1}\rangle^{(1)}$ and~$\langle J^{2}\rangle^{(2)}$ results from a cancellation 
between the contributions of the two terms on the right-hand side of eq.~\eqref{eq:last_eq_pert_th}.
As for~$\langle J^{3}\rangle^{(1)}$, we find 
a nontrivial cancellation of all finite temperature contributions proportional to~$\mathcal{T}_{B}$
\begin{align}
\langle J^{3}\rangle^{(1)}&={\mathcal{T}_{B}-B\over 32\pi^{2}}\Lambda^{2}
+{B\over 24}\left(T_{0}^{2}+{3\over \pi^{2}}\mu_{L}^{2}\right).
\label{eq:kzgen_J3(1)final}
\end{align}
This absence of thermal linear torsional corrections
agrees with 
the general expressions~\eqref{eq:J_0_general} and~\eqref{eq:J^i_general}, 
given that~$\mu_{S}=0$ for the torsion tensor shown in
eq.~\eqref{eq:kzgen_magtorsion}.

Let us focus for a moment in the cutoff-dependent vacuum terms in~\eqref{eq:kzgen_J3(1)final}. 
In our perturbative computation, the
coincidence limit~$\boldsymbol{r}'\rightarrow \boldsymbol{r}$ 
in~\eqref{eq:last_eq_pert_th} has been taken before carrying out the integration over 
proper time coming from the zeroth-order Green function~\eqref{eq:GF_zero_order}.
However, would we have performed this integration prior to the coincidence limit, the torsional
term in the vacuum contribution would be regularized to zero, with only the part scaling
as~$B\Lambda^{2}$ remaining. The finite temperature terms, on the other hand, are
not affected by the order ambiguity. It should also be mentioned that,
taking the coincidence limit first, the presence of a
term proportional to~$\mathcal{T}_{B}\Lambda^{2}$ in~\eqref{eq:kzgen_J3(1)final}
agrees with the general form of the vacuum chiral
current obtained in ref.~\cite{Valle:2021nfv} using the 
method of descent.

To be more precise, the origin of the ambiguity traces back to the integral
\begin{align}
I(\boldsymbol{r},\boldsymbol{r}')
=\int_{0}^{\infty}{ds\over s^{5/2}}\big[(\boldsymbol{r}-\boldsymbol{r}')^{2}-2s-4(\omega_{n}-i\mu_{L})^{2}s^{2}\big]
e^{-s(\omega_{n}-i\mu)^{2}-{1\over 4s}(\boldsymbol{r}-\boldsymbol{r}')^{2}},
\label{eq:amb_iden_type}
\end{align}
which is convergent and 
vanishes when evaluated for~$\boldsymbol{r}'\neq\boldsymbol{r}$. 
Taking the coincidence limit first, on the other hand, makes the proper time integration ill
defined and forces 
the introduction of a UV~cutoff. After inserting the sums over Matsubara frequencies, the proper time
integration is carried out to give
\begin{align}
-2\int_{\Lambda^{-2}}^{\infty}{ds\over s^{3/2}}\big[1+2(\omega_{n}-i\mu_{L})^{2}s\big]
\sum_{n\in\mathbb{z}}e^{-s(\omega_{n}-i\mu_{L})^{2}}=-{2\beta_{0}\over\sqrt{\pi}}\Lambda^{2}
+\mathcal{O}(e^{-\beta_{0}^{2}\Lambda^{2}}),
\label{eq:coinc_limit_2ninit}
\end{align}
with no power-like contributions in either the temperature or the chemical potential. 
Recasting the sums in terms of
Jacobi theta functions, it can be seen that~\eqref{eq:coinc_limit_2ninit} follows again from applying
the identity~\eqref{eq:magic_identity} with~$n=2$.

An explicit evaluation of the energy momentum tensor components shows 
the absence of linear torsional corrections, confirming
the results found in sec.~\ref{sec:general_analysis} for a background 
with~$T^{i}_{\,\,\,jk}=T^{i}_{\,\,\,k0}=0$. Only the heat current 
\begin{align}
q^{3}={B\over 12}\mu_{L}
\left(T_{0}^{2}+{1\over \pi^{2}}\mu_{L}^{2}\right),
\end{align} 
picks up a linear dependence on the vortical parameter~$B$. 
The finite temperature contributions to the spin energy potential linear in the background data
show no dependence on~$\mathcal{T}_{B}$ either. Indeed, the only nonzero components 
are
\begin{align}
\langle\Psi^{012}\rangle^{(1)}={B\over 96}\left(T_{0}^{2}+{3\over \pi^{2}}\mu_{L}^{2}\right),
\end{align}
together with the ones obtained by permutation of the three antisymmetric indices.

To close this section, let us point out that the analysis  
presented above can be easily generalized to
the geometry defined by the tetrad
\begin{align}
e^{a}=\left(dt-\omega_{ij}x^{i}dx^{j},dx^{1},dx^{2},dx^{3}\right),
\end{align} 
with spin connection
\begin{align}
\omega_{0i}&=-\omega_{i0}={1\over 2}(T^{0}_{\,\,\,ij}+2\omega_{ij})e^{j}, 
\nonumber \\[0.2cm]
\omega_{ij}&=-\omega_{ji}=-{1\over 2}(T^{0}_{\,\,\,ij}+2\omega_{ij})e^{0},
\end{align}
for constant~$\omega_{ij}$ and~$T^{0}_{\,\,\,ij}$, both antisymmetric in their lower indices. 
This gives nonzero values for the vorticity and torsion two-forms
\begin{align}
\omega&={1\over 2}\omega_{ij}dx^{i}dx^{j}, \nonumber \\[0.2cm]
T^{a}&={1\over 2}\delta^{a}_{0}T^{0}_{\,\,\,ij}dx^{i}dx^{j},
\end{align} 
where from the latter identity we conclude that~$\mu_{S}=0$.
A calculation of the thermal constitutive relations renders the results
\begin{align}
\langle J^{i}\rangle^{(1)}&=-{1\over 12}
\left(T_{0}^{2}+{3\over \pi^{2}}\mu_{L}^{2}\right)\omega^{i}, \nonumber \\[0.2cm]
q^{i}&=-{1\over 6}\mu_{L}
\left(T_{0}^{2}+{1\over \pi^{2}}\mu_{L}^{2}\right)\omega^{i}, \\[0.2cm]
\langle \Psi^{0ij}\rangle^{(1)}&=-{1\over 48}
\left(T_{0}^{2}+{3\over \pi^{2}}\mu_{L}^{2}\right)\omega^{ij}.
\nonumber
\end{align}
Here we have introduced
the vorticity vector~$\omega^{i}={1\over 2}\epsilon^{ijk}\omega_{jk}$ and 
dropped vacuum contributions. The well-known results for the chiral vortical effect are thus retrieved, without any
role being played by~$T^{0}_{\,\,\,ij}\neq 0$.

\section{A background with Nieh-Yan anomaly}
\label{sec:NY_anomaly}

A common feature of the two geometries studied so far is that their curvature two-forms are
of second order in the background data. One consequence of this 
is that their corresponding Nieh-Yan invariants vanish
at linear order, and no torsional
violations in the conservation of the axial current are found\footnote{In the case of the background
studied in sec.~\ref{sec:general_analysis}, the explicit calculation
of the divergence of the axial-vector current carried out in refs.~\cite{Huang:2019haq,Laurila:2020yll} 
shows that the anomaly is indeed
quadratic in the background data.}. 
To avoid this state of affairs, let us consider the background defined by the
trivial tetrad
\begin{align}
e^{a}=\big(dt,dx,dy,dz\big),
\label{eq:trivial_background}
\end{align}
together with a spin connection one-form describing a nonzero spin chemical potential with
\begin{align}
\mu_{12}=-\mu_{21}={\mathcal{T}_{E}\over 2}z,
\label{eq:NY_chempot}
\end{align}
and vanishing magnetic components,~$\boldsymbol{\omega}^{a}_{\,\,\,b}=0$ [see the second identity in~\eqref{eq:spin_connection_EMdecompgen}]. The 
associated torsion two-form is found to be purely electric
\begin{align}
T^{a}={\mathcal{T}_{E}\over 2}dt\big(\delta^{a}_{1}zdy-\delta^{a}_{2}zdx\big),
\label{eq:el_torsion_NY}
\end{align}
while the nonzero components of the curvature
\begin{align}
R_{12}=-R_{21}=-{\mathcal{T}_{E}\over 2}e^{0}e^{3},
\label{eq:curvature_NY}
\end{align}
are linear in the single background datum~$\mathcal{T}_{E}$. 
Since~$u=-dt$, this geometry can be used to describes a chiral fermion
fluid coupled to torsion with no vorticity or acceleration, for which the
equilibrium constraint~\eqref{eq:equilibrium_constr} is trivially satisfied.

The one-particle Green thermal equation~\eqref{eq:GF_def_eom_frequencies1} for the Weyl
Hamiltonian
\begin{align}
\mathcal{H}=i\sigma_{x}\partial_{x}+i\sigma_{y}\partial_{y}+i\sigma_{z}\partial_{z}
-\mu_{L}-{1\over 4}\mathcal{T}_{E}z\sigma_{z},
\label{eq:hamiltonian_NYferm}
\end{align}
can be solved in perturbation theory with respect to~$\mathcal{T}_{E}$, as we did in the case discussed in
sec.~\ref{sec:T0ij}. For the chiral current, we find the following results for the 
linear torsional terms
\begin{align}
\langle J_{0}\rangle^{(1)}&=\langle J^{1}\rangle^{(1)}=\langle J^{(2)}\rangle^{(1)}=0, \nonumber \\[0.2cm]
\langle J^{3}\rangle^{(1)}&={\Lambda^{2}\over 32\pi^{2}}\mathcal{T}_{E}z,
\label{eq:NY_current}
\end{align} 
where~$\Lambda$ is the proper time cutoff. Unlike the background studied in the
previous section, here there are
no finite-temperature linear torsional corrections, again as a consequence of the identity~\eqref{eq:magic_identity} with~$n=2$.
We now take the divergence of the computed current to obtain a nonzero value for the 
Nieh-Yan anomaly (cf.~\cite{Valle:2021nfv})
\begin{align}
\partial_{\mu}\langle J^{\mu}\rangle&={\Lambda^{2}\over 32\pi^{2}}\mathcal{T}_{E}
\nonumber \\[0.2cm]
&={\Lambda^{2}\over 32\pi^{2}}
\star\big(T_{a}T^{a}-e_{a}R^{a}_{\,\,\,b}e^{b}\big),
\label{eq:NY_anomaly_backgrd}
\end{align}
where the Nieh-Yan invariant in the second line
is computed to linear order in~$\mathcal{T}_{E}$ and 
the star denotes the four-dimensional Hodge dual. These results are totally consistent
with the general structure of the vacuum chiral current found in~\cite{Valle:2021nfv}.
It should be stressed as well that 
this absence of linear thermal corrections in both the chiral current and 
the Nieh-Yan anomaly
provide a counterexample to the conjecture stated in~\cite{Nissinen:2019wmh}, according to which 
the coefficient of the Nieh-Yan anomaly
should receive corrections proportional to~$T_{0}^{2}$.

As in the case of the background studied in the 
previous section, here as well the presence of torsional vacuum terms in~\eqref{eq:NY_current} depends
on the order in which the coincidence limit and the proper time integration are carried out. Integrating
first in proper time when evaluating the 
Green function in perturbation theory, we find that the vacuum contribution vanishes after the coincidence limit
is properly taken, and as a consequence the Nieh-Yan anomaly is also regularized to zero. 
The ambiguity once again stems from an integral of the type shown in eq.~\eqref{eq:amb_iden_type}. This suggests that,
at least in this case, the Nieh-Yan anomaly could be regarded as an artifact of the 
regularization\footnote{For a discussion of the Nieh-Yan anomaly vs. the standard 
Adler-Bell-Jackiw anomaly,
see ref.~\cite{Soo:1998ev}.}.

Concerning the energy-momentum tensor, only the heat current acquires a term
proportional to~$\mathcal{T}_{E}$
\begin{align}
q^{3}=-{\mathcal{T}_{E}\over 48}z\mu_{L}\left( T_{0}^{2}+{1\over \pi^{2}}\mu_{L}^{2}\right),
\end{align}
with~$q^{0}=q^{1}=q^{2}=0$. Once more, the results for
both the chiral current and the energy-momentum tensor obtained from the microscopic calculation 
match with the expressions derived in sec.~\ref{sec:general_analysis} applied to a background 
with~$\sigma=0=a_{i}$,~$g_{ij}=\delta_{ij}$, and a torsion tensor whose
only nonzero components are~$T^{1}_{\,\,\,02}=-T^{1}_{\,\,\,20}=T^{2}_{\,\,\,01}
=-T^{2}_{\,\,\,10}={1\over 2}\mathcal{T}_{E}$.

\section{Closing remarks}
\label{sec:conclusions}

In this paper we have studied finite 
temperature effects in chiral fluids at equilibrium coupled to background torsion.
At the level of the equilibrium partition function,
all dependence on the torsion at 
first order in the derivative expansion enters through a shift in the chiral chemical potential. 
Our main result is a full determination of the general structure of
the linear torsional corrections to the 
constitutive relations of the chiral current, energy-momentum tensor, 
and spin energy potential. One interesting conclusion to be extracted from these results
is the existence of nonzero
torsion configurations which do not 
leave imprints in the constitutive relations of these three conserved currents.  
Another important upshot of our analysis is that the constitutive relation of the 
energy-momentum tensor contains nondissipative torsional
terms analog to the $(2+1)$-dimensional Hall viscosity. 

We also carried out a first-principle, microscopic computation of these 
constitutive relations for some particular 
geometries, confirming in all 
cases the general results obtained in sec.~\ref{sec:general_analysis}.
Unlike most of the torsional backgrounds with potential
applications to condensed matter physics studied in the literature
(see, for
example,~\cite{Hughes:2012vg,Parrikar:2014usa,Khaidukov:2018oat,Huang:2019haq,
Nissinen:2019mkw,Huang:2020ypv,Laurila:2020yll}), here we have focused our attention on geometries
with nonzero spin connection and curvature. This not only opens a way of 
adding curvature and thus implementing 
the effects of disclinations as well as dislocations, but also allows the disentanglement of  
torsion from the background data parametrizing other fluid properties. 
This feature might in fact shed light on the interplay between torsion and vorticity, 
a subject of some recent controversy
(see, for example,~\cite{Ferreiros:2020uda,Chernodub:2021nff,Nissinen:2021gke}).

The geometrical background studied in sec.~\ref{sec:T0ij} offers a good test bench for this
task. As explained there, the model includes both vorticity and torsion,
parametrized by the two {\em independent} background 
data~$B$ and~$\mathcal{T}_{B}$ (or $\omega_{ij}$ and~$T^{0}_{\,\,\,ij}$,
in the generalization discussed at the section's end),
their difference (resp.,~$T^{0}_{\,\,\,ij}+2\omega_{ij}$) determining the 
spin connection and the curvature.
At linear order, however, thermal corrections to the constitutive relations 
only depend on vorticity, thus indicating
the absence of genuine torsional transport effects.
The particular case~$B=\mathcal{T}_{B}$ 
(resp.,~$\omega_{ij}=-{1\over 2}T^{0}_{\,\,\,ij}$) deserves closer attention. 
The torsional background in this instance reduces to the one
used in ref.~\cite{Khaidukov:2018oat}, whose results we generalized to 
allow for a nonzero chiral chemical potential. 
Since~$\omega^{a}_{\,\,\,b}=0=\mathfrak{a}$,
torsion can now be
interpreted as a particular way of implementing vorticity, based on the
identification
\begin{align}
\omega={1\over 2}du=-{1\over 2}de^{0}=-{1\over 2}T^{0}.
\label{eq:identification} 
\end{align}
Let us point out however that this identity 
is not Lorentz covariant, the left-hand side being a scalar
while the right-hand side is the time component of a Lorentz vector, so such a direct link
between torsion and vorticity is problematic in a generic geometry. This indicates that
the identification of vorticity with torsion only works in very particular cases.

Our microscopic analyses also bring forward an interesting
feature of the vacuum contributions to the chiral current, with some bearing on 
the results of ref.~\cite{Valle:2021nfv}. We have seen how the torsional vacuum terms in 
the backgrounds studied in secs.~\ref{sec:T0ij} 
and~\ref{sec:NY_anomaly}, including the Nieh-Yan anomaly 
in the second case, can be removed by
an appropriate prescription in the order in which the integration over proper time and
the coincidence limit are taken. It is important to stress that
this ambiguity only affects the torsional part
of the vacuum contribution, whereas both 
finite temperature contributions and the 
vacuum terms proportional to other background data remain unchanged. 

This last
trait is also manifest in the background of secs.~\ref{sec:microscopic}, where it can 
be shown that the vacuum term 
proportional to~$B\Lambda^{2}$ is free from order ambiguities. The distinct 
feature of this case is that all torsion dependence in the 
Green function exclusively enters through a nonzero value of
the torsional chemical potential~$\mu_{S}$, whereas for the other two 
backgrounds studied here this quantity is zero and torsion dependence comes from the spatial components 
of the torsional gauge field. 
This is why the corresponding finite temperature
constitutive relations of the chiral current display no torsional linear corrections, as implied also by the results
of sec.~\ref{sec:general_analysis}. 
Notice, however, that choosing the prescription preserving the torsional terms in the 
current cutoff-dependent piece leads to results agreeing with the general form of the 
zero temperature current derived in~\cite{Valle:2021nfv}.

Finally, although this work  is not  directly concerned with the nature of the Nieh-Yan anomaly, 
it is worth-mentioning that the ambiguity we have found related to the order in which the 
integration over proper time and the coincidence limit are carried out seems to indicate that this anomaly 
could be avoided, at least in some cases.  
Anyhow, on the broad issue of the physical role of the Nieh-Yan anomaly 
the jury is still out (see, for example, 
refs.~\cite{Chandia:1997hu,Soo:1998ev,
Huang:2019haq,Nissinen:2019kld,Nissinen:2019wmh,Nissinen:2019mkw,Huang:2019adx,Huang:2020ypv,
Imaki:2020csc,Liu:2021bic,Nissinen:2021gke,Chernodub:2021nff,Valle:2021nfv,Rasulian:2023gtb}).

\acknowledgments

We thank Carlos Hoyos for interesting discussions.
This work has been supported by 
Spanish Science Ministry grants PID2021-123703NB-C21
(MCIU/AEI/FEDER, EU) and PID2021-123703NB-C22
(MCIU/AEI/FEDER, EU), as well as by Basque Government
grant IT1628-22. 

\appendix
\section{A derivation of the identity~\eqref{eq:magic_identity}}
\label{app:magic_identity}

Let us consider the integral
\begin{align}
I=\int_{0}^{\Lambda^{2}}dx\,x^{n-2}\theta_{4}(z|i\alpha x),
\label{eq:magic_identity_general_app}
\end{align}
with~$n>1$ and~$\alpha$ a real positive number, and rewrite it as
\begin{align}
I&={1\over n-1}\int_{0}^{\Lambda^{2}}dx\,{d\over dx}(x^{n-1})\theta_{4}(z|i\alpha x)
\nonumber \\[0.2cm]
&={1\over n-1}\Lambda^{2n-2}-{1\over n-1}\int_{0}^{\Lambda^{2}}dx\,x^{n-1}{\partial\over\partial x}
\theta_{4}(z|i\alpha x)+\mathcal{O}(e^{-\alpha\Lambda^{2}}).
\label{eq:integral_intermediate1}
\end{align}
In the second line an integration by parts has been performed and
all terms of order~$e^{-\alpha\Lambda^{2}}$ and smaller have been dropped.
Next, we
implement the
heat equation satisfied by the Jacobi theta function~\cite{ErdelyiVol2}
\begin{align}
{\partial\over\partial x}\theta_{4}(z|i\alpha x)&={\alpha\over 4\pi}\theta_{4}''(z|i\alpha x),
\end{align}
so eq.~\eqref{eq:integral_intermediate1} is recast as
\begin{align}
I&={1\over n-1}\Lambda^{2n-2}-{\alpha \over 4(n-1)\pi}
\int_{0}^{\Lambda^{2}}dx\,x^{n-1}
\theta_{4}''(z|i\alpha x)+\mathcal{O}(e^{-\alpha\Lambda^{2}}).
\end{align}
Identifying this expression with the original integral in~\eqref{eq:magic_identity_general_app}, we 
finally arrive at the identity
\begin{align}
\int_{0}^{\Lambda^{2}}dx\,x^{n-2}\theta_{4}(z|i\alpha x)
+{\alpha \over 4(n-1)\pi}
&\int_{0}^{\infty}dx\,x^{n-1}
\theta_{4}''(z|i\alpha x) \nonumber \\[0.2cm]
&={1\over n-1}\Lambda^{2n-2}
+\mathcal{O}(e^{-\alpha\Lambda^{2}}),
\label{eq:magic_identity_generalex}
\end{align}
where the upper limit in the second integration was extended all the way to infinity, since
the absence of a zero mode in the Jacobi theta function makes the integral convergent.
After changing variables to~$s=x^{-1}$,
and setting~$z={\mu\beta_{0}\over 2\pi i}$ 
and~$\alpha={\beta_{0}^{2}\over 4\pi}$,
we retrieve eq.~\eqref{eq:magic_identity}. Incidentally, by differentiating 
eq.~\eqref{eq:magic_identity_generalex} with respect to~$z$, a sequence of identities relating convergent
integrals involving the $k$th and $(k+2)$th derivatives of the Jacobi theta function are obtained
\begin{align}
\int_{0}^{\infty}dx\,x^{n-2}\theta_{4}^{(k)}(z|i\alpha x)
=-{\alpha \over 4(n-1)\pi}
&\int_{0}^{\infty}dx\,x^{n-1}
\theta_{4}^{(k+2)}(z|i\alpha x),
\end{align} 
with~$k>1$.

\bibliographystyle{JHEP}
\bibliography{biblio_file}

\providecommand{\href}[2]{#2}\begingroup\raggedright\begin{thebibliography}{10}

\bibitem{Hehl:1976kj}
F.~Hehl, P.~von~der Heyde, G.~Kerlick and J.~Nester, \emph{{General Relativity
  with Spin and Torsion: Foundations and Prospects}},
  \href{https://doi.org/10.1103/RevModPhys.48.393}{\emph{Rev. Mod. Phys.}
  {\bfseries 48} (1976) 393}.

\bibitem{Shapiro:2001rz}
I.~Shapiro, \emph{{Physical aspects of the space-time torsion}},
  \href{https://doi.org/10.1016/S0370-1573(01)00030-8}{\emph{Phys. Rept.}
  {\bfseries 357} (2002) 113}
  [\href{https://arxiv.org/abs/hep-th/0103093}{{\ttfamily hep-th/0103093}}].

\bibitem{Kondo1952}
K.~Kondo, \emph{{On the geometrical and physical foundations of the theory of
  yielding}},  in \emph{{``Proceedings of the 2nd Japan National Congress for
  Applied Mechanics'', {\rm Tokyo}}}, pp.~41--47, 1952.

\bibitem{Bilby:1955}
B.~A. Bilby, R.~Bullough and E.~Smith, \emph{{Continuous Distributions of
  Dislocations: A New Application of the Methods of Non-Riemannian Geometry}},
  {\emph{Proc. R. Soc. Lond. A} {\bfseries 231} (1955) 263}.

\bibitem{Katanaev:1992kh}
M.~O. Katanaev and I.~V. Volovich, \emph{{Theory of defects in solids and
  three-dimensional gravity}},
  \href{https://doi.org/10.1016/0003-4916(52)90040-7}{\emph{Annals Phys.}
  {\bfseries 216} (1992) 1}.

\bibitem{Hehl:2007bn}
F.~W. Hehl and Y.~N. Obukhov, \emph{{Élie Cartan's torsion in geometry and in
  field theory, an essay}}, {\emph{Annales Fond. Broglie} {\bfseries 32} (2007)
  157} [\href{https://arxiv.org/abs/0711.1535}{{\ttfamily 0711.1535}}].

\bibitem{Kleman:2008zz}
M.~Kleman and J.~Friedel, \emph{{Disclinations, dislocations, and continuous
  defects: A reappraisal}},
  \href{https://doi.org/10.1103/RevModPhys.80.61}{\emph{Rev. Mod. Phys.}
  {\bfseries 80} (2008) 61} [\href{https://arxiv.org/abs/0704.3055}{{\ttfamily
  0704.3055}}].

\bibitem{Katanaev:2021bje}
M.~O. Katanaev, \emph{{Disclinations in the geometric theory of defects}},
  \href{https://doi.org/10.1134/S0081543821020097}{\emph{Proc. Steklov Inst.
  Math.} {\bfseries 313} (2021) 1}
  [\href{https://arxiv.org/abs/2108.07177}{{\ttfamily 2108.07177}}].

\bibitem{Hughes:2012vg}
T.~L. Hughes, R.~G. Leigh and O.~Parrikar, \emph{{Torsional Anomalies, Hall
  Viscosity, and Bulk-boundary Correspondence in Topological States}},
  \href{https://doi.org/10.1103/PhysRevD.88.025040}{\emph{Phys. Rev. D}
  {\bfseries 88} (2013) 025040}
  [\href{https://arxiv.org/abs/1211.6442}{{\ttfamily 1211.6442}}].

\bibitem{Parrikar:2014usa}
O.~Parrikar, T.~L. Hughes and R.~G. Leigh, \emph{{Torsion, Parity-odd Response
  and Anomalies in Topological States}},
  \href{https://doi.org/10.1103/PhysRevD.90.105004}{\emph{Phys. Rev. D}
  {\bfseries 90} (2014) 105004}
  [\href{https://arxiv.org/abs/1407.7043}{{\ttfamily 1407.7043}}].

\bibitem{Khaidukov:2018oat}
Z.~Khaidukov and M.~Zubkov, \emph{{Chiral torsional effect}},
  \href{https://doi.org/10.1134/S0021364018220046}{\emph{JETP Lett.} {\bfseries
  108} (2018) 670} [\href{https://arxiv.org/abs/1812.00970}{{\ttfamily
  1812.00970}}].

\bibitem{Nissinen:2019kld}
J.~Nissinen, \emph{{Emergent spacetime and gravitational Nieh-Yan anomaly in
  chiral $p+ip$ Weyl superfluids and superconductors}},
  \href{https://doi.org/10.1103/PhysRevLett.124.117002}{\emph{Phys. Rev. Lett.}
  {\bfseries 124} (2020) 117002}
  [\href{https://arxiv.org/abs/1909.05846}{{\ttfamily 1909.05846}}].

\bibitem{Huang:2019haq}
Z.-M. Huang, B.~Han and M.~Stone, \emph{{Nieh-Yan anomaly: Torsional Landau
  levels, central charge, and anomalous thermal Hall effect}},
  \href{https://doi.org/10.1103/PhysRevB.101.125201}{\emph{Phys. Rev. B}
  {\bfseries 101} (2020) 125201}
  [\href{https://arxiv.org/abs/1911.00174}{{\ttfamily 1911.00174}}].

\bibitem{Nissinen:2019wmh}
J.~Nissinen and G.~E. Volovik, \emph{{On thermal Nieh\textendash{}Yan anomaly
  in topological Weyl materials}},
  \href{https://doi.org/10.1134/S0021364019240020}{\emph{Pisma Zh. Eksp. Teor.
  Fiz.} {\bfseries 110} (2019) 797}
  [\href{https://arxiv.org/abs/1911.03382}{{\ttfamily 1911.03382}}].

\bibitem{Nissinen:2019mkw}
J.~Nissinen and G.~Volovik, \emph{{Thermal Nieh-Yan anomaly in Weyl
  superfluids}},
  \href{https://doi.org/10.1103/PhysRevResearch.2.033269}{\emph{Phys. Rev.
  Res.} {\bfseries 2} (2020) 033269}
  [\href{https://arxiv.org/abs/1909.08936}{{\ttfamily 1909.08936}}].

\bibitem{Huang:2019adx}
Z.-M. Huang, B.~Han and M.~Stone, \emph{{Hamiltonian approach to the torsional
  anomalies and its dimensional ladder}},
  \href{https://doi.org/10.1103/PhysRevB.101.165201}{\emph{Phys. Rev. B}
  {\bfseries 101} (2020) 165201}
  [\href{https://arxiv.org/abs/1912.06051}{{\ttfamily 1912.06051}}].

\bibitem{Huang:2020ypv}
Z.-M. Huang and B.~Han, \emph{{Torsional Anomalies and Bulk-Dislocation
  Correspondence in Weyl Systems}},
  \href{https://arxiv.org/abs/2003.04853}{{\ttfamily 2003.04853}}.

\bibitem{Imaki:2020csc}
S.~Imaki and Z.~Qiu, \emph{{Chiral torsional effect with finite temperature,
  density and curvature}},
  \href{https://doi.org/10.1103/PhysRevD.102.016001}{\emph{Phys. Rev. D}
  {\bfseries 102} (2020) 016001}
  [\href{https://arxiv.org/abs/2004.11899}{{\ttfamily 2004.11899}}].

\bibitem{Ferreiros:2020uda}
Y.~Ferreiros and K.~Landsteiner, \emph{{On chiral responses to geometric
  torsion}}, \href{https://doi.org/10.1016/j.physletb.2021.136419}{\emph{Phys.
  Lett. B} {\bfseries 819} (2021) 136419}
  [\href{https://arxiv.org/abs/2011.10535}{{\ttfamily 2011.10535}}].

\bibitem{Manes:2020zdd}
J.~L. Ma\~nes, M.~Valle and M.~\'A. V\'azquez-Mozo, \emph{{Chiral torsional
  effects in anomalous fluids in thermal equilibrium}},
  \href{https://doi.org/10.1007/JHEP05(2021)209}{\emph{JHEP} {\bfseries 05}
  (2021) 209} [\href{https://arxiv.org/abs/2012.08449}{{\ttfamily
  2012.08449}}].

\bibitem{Liu:2021bic}
C.-X. Liu, \emph{{Probing Nieh-Yan anomaly through phonon dynamics in the
  Kramers-Weyl semimetals of chiral crystals}},
  \href{https://doi.org/10.1103/PhysRevB.106.115102}{\emph{Phys. Rev. B}
  {\bfseries 106} (2022) 115102}
  [\href{https://arxiv.org/abs/2104.04859}{{\ttfamily 2104.04859}}].

\bibitem{Chernodub:2021nff}
M.~N. Chernodub, Y.~Ferreiros, A.~G. Grushin, K.~Landsteiner and M.~A.~H.
  Vozmediano, \emph{{Thermal transport, geometry, and anomalies}},
  \href{https://doi.org/10.1016/j.physrep.2022.06.002}{\emph{Phys. Rept.}
  {\bfseries 977} (2022) 1} [\href{https://arxiv.org/abs/2110.05471}{{\ttfamily
  2110.05471}}].

\bibitem{Nissinen:2021gke}
J.~Nissinen and G.~E. Volovik, \emph{{Anomalous chiral transport with vorticity
  and torsion: Cancellation of two mixed gravitational anomaly currents in
  rotating chiral p+ip Weyl condensates}},
  \href{https://doi.org/10.1103/PhysRevD.106.045022}{\emph{Phys. Rev. D}
  {\bfseries 106} (2022) 045022}
  [\href{https://arxiv.org/abs/2111.08639}{{\ttfamily 2111.08639}}].

\bibitem{Valle:2021nfv}
M.~Valle and M.~\'A. V\'azquez-Mozo, \emph{{On Nieh-Yan transport}},
  \href{https://doi.org/10.1007/JHEP03(2022)177}{\emph{JHEP} {\bfseries 03}
  (2022) 177} [\href{https://arxiv.org/abs/2112.02003}{{\ttfamily
  2112.02003}}].

\bibitem{Amitani:2022xev}
T.~Amitani and Y.~Nishida, \emph{{Torsion-induced chiral magnetic current in
  equilibrium}}, \href{https://doi.org/10.1016/j.aop.2022.169181}{\emph{Annals
  Phys.} {\bfseries 448} (2023) 169181}
  [\href{https://arxiv.org/abs/2204.13415}{{\ttfamily 2204.13415}}].

\bibitem{Hidaka:2012rj}
Y.~Hidaka, Y.~Hirono, T.~Kimura and Y.~Minami,
  \emph{{Viscoelastic-electromagnetism and Hall viscosity}},
  \href{https://doi.org/10.1093/ptep/pts063}{\emph{PTEP} {\bfseries 2013}
  (2013) 013A02} [\href{https://arxiv.org/abs/1206.0734}{{\ttfamily
  1206.0734}}].

\bibitem{Gromov:2014vla}
A.~Gromov and A.~G. Abanov, \emph{{Thermal Hall Effect and Geometry with
  Torsion}}, \href{https://doi.org/10.1103/PhysRevLett.114.016802}{\emph{Phys.
  Rev. Lett.} {\bfseries 114} (2015) 016802}
  [\href{https://arxiv.org/abs/1407.2908}{{\ttfamily 1407.2908}}].

\bibitem{Geracie:2014mta}
M.~Geracie, S.~Golkar and M.~M. Roberts, \emph{{Hall viscosity, spin density,
  and torsion}},  \href{https://arxiv.org/abs/1410.2574}{{\ttfamily
  1410.2574}}.

\bibitem{Sumiyoshi:2015eda}
H.~Sumiyoshi and S.~Fujimoto, \emph{{Torsional Chiral Magnetic Effect in a Weyl
  Semimetal with a Topological Defect}},
  \href{https://doi.org/10.1103/PhysRevLett.116.166601}{\emph{Phys. Rev. Lett.}
  {\bfseries 116} (2016) 166601}
  [\href{https://arxiv.org/abs/1509.03981}{{\ttfamily 1509.03981}}].

\bibitem{Laurila:2020yll}
S.~Laurila and J.~Nissinen, \emph{{Torsional Landau levels and geometric
  anomalies in condensed matter Weyl systems}},
  \href{https://doi.org/10.1103/PhysRevB.102.235163}{\emph{Phys. Rev. B}
  {\bfseries 102} (2020) 235163}
  [\href{https://arxiv.org/abs/2007.10682}{{\ttfamily 2007.10682}}].

\bibitem{Valle:2015hfa}
M.~Valle, \emph{{Torsional response of relativistic fermions in $2+1$
  dimensions}}, \href{https://doi.org/10.1007/JHEP07(2015)006}{\emph{JHEP}
  {\bfseries 07} (2015) 006}
  [\href{https://arxiv.org/abs/1503.04020}{{\ttfamily 1503.04020}}].

\bibitem{Huang:2021luf}
Z.-M. Huang, B.~Han and X.-Q. Sun, \emph{{Torsion, energy magnetization, and
  thermal Hall effect}},
  \href{https://doi.org/10.1103/PhysRevB.105.085104}{\emph{Phys. Rev. B}
  {\bfseries 105} (2022) 085104}
  [\href{https://arxiv.org/abs/2105.01600}{{\ttfamily 2105.01600}}].

\bibitem{Nissinen:2023bgl}
J.~Nissinen, \emph{{Emergent geometry, torsion and anomalies in
  non-relativistic topological matter}},
  \href{https://doi.org/10.1088/1742-6596/2531/1/012002}{\emph{J. Phys. Conf.
  Ser.} {\bfseries 2531} (2023) 012002}.

\bibitem{Gallegos:2020otk}
A.~D. Gallegos and U.~G\"ursoy, \emph{{Holographic spin liquids and Lovelock
  Chern-Simons gravity}},
  \href{https://doi.org/10.1007/JHEP11(2020)151}{\emph{JHEP} {\bfseries 11}
  (2020) 151} [\href{https://arxiv.org/abs/2004.05148}{{\ttfamily
  2004.05148}}].

\bibitem{Gallegos:2021bzp}
A.~D. Gallegos, U.~G\"ursoy and A.~Yarom, \emph{{Hydrodynamics of spin
  currents}},
  \href{https://doi.org/10.21468/SciPostPhys.11.2.041}{\emph{SciPost Phys.}
  {\bfseries 11} (2021) 041}
  [\href{https://arxiv.org/abs/2101.04759}{{\ttfamily 2101.04759}}].

\bibitem{Hongo:2021ona}
M.~Hongo, X.-G. Huang, M.~Kaminski, M.~Stephanov and H.-U. Yee,
  \emph{{Relativistic spin hydrodynamics with torsion and linear response
  theory for spin relaxation}},
  \href{https://arxiv.org/abs/2107.14231}{{\ttfamily 2107.14231}}.

\bibitem{Gallegos:2022jow}
A.~D. Gallegos, U.~Gursoy and A.~Yarom, \emph{{Hydrodynamics, spin currents and
  torsion}}, \href{https://doi.org/10.1007/JHEP05(2023)139}{\emph{JHEP}
  {\bfseries 05} (2023) 139}
  [\href{https://arxiv.org/abs/2203.05044}{{\ttfamily 2203.05044}}].

\bibitem{Florkowski:2018fap}
W.~Florkowski, A.~Kumar and R.~Ryblewski, \emph{{Relativistic hydrodynamics for
  spin-polarized fluids}},
  \href{https://doi.org/10.1016/j.ppnp.2019.07.001}{\emph{Prog. Part. Nucl.
  Phys.} {\bfseries 108} (2019) 103709}
  [\href{https://arxiv.org/abs/1811.04409}{{\ttfamily 1811.04409}}].

\bibitem{Bhadury:2021oat}
S.~Bhadury, J.~Bhatt, A.~Jaiswal and A.~Kumar, \emph{{New developments in
  relativistic fluid dynamics with spin}},
  \href{https://doi.org/10.1140/epjs/s11734-021-00020-4}{\emph{Eur. Phys. J.
  ST} {\bfseries 230} (2021) 655}
  [\href{https://arxiv.org/abs/2101.11964}{{\ttfamily 2101.11964}}].

\bibitem{Hattori:2022hyo}
K.~Hattori, M.~Hongo and X.-G. Huang, \emph{{New Developments in Relativistic
  Magnetohydrodynamics}},
  \href{https://doi.org/10.3390/sym14091851}{\emph{Symmetry} {\bfseries 14}
  (2022) 1851} [\href{https://arxiv.org/abs/2207.12794}{{\ttfamily
  2207.12794}}].

\bibitem{Chu:2022bhj}
C.-S. Chu and R.-X. Miao, \emph{{Chiral current induced by torsional Weyl
  anomaly}}, \href{https://doi.org/10.1103/PhysRevB.107.205410}{\emph{Phys.
  Rev. B} {\bfseries 107} (2023) 205410}
  [\href{https://arxiv.org/abs/2210.01382}{{\ttfamily 2210.01382}}].

\bibitem{Hehl:1971qi}
F.~W. Hehl and B.~K. Datta, \emph{{Nonlinear spinor equation and asymmetric
  connection in general relativity}},
  \href{https://doi.org/10.1063/1.1665738}{\emph{J. Math. Phys.} {\bfseries 12}
  (1971) 1334}.

\bibitem{Datta:1971id}
B.~K. Datta, \emph{{Spinor fields in general relativity. 2. generalized field
  equations and application to the dirac field}},
  \href{https://doi.org/10.1007/BF02738159}{\emph{Nuovo Cim. B} {\bfseries 6}
  (1971) 16}.

\bibitem{Freedman:2012zz}
D.~Z. Freedman and A.~Van~Proeyen, \emph{{Supergravity}}. Cambridge University
  Press, 2012.

\bibitem{Banerjee:2012iz}
N.~Banerjee, J.~Bhattacharya, S.~Bhattacharyya, S.~Jain, S.~Minwalla and
  T.~Sharma, \emph{{Constraints on Fluid Dynamics from Equilibrium Partition
  Functions}}, \href{https://doi.org/10.1007/JHEP09(2012)046}{\emph{JHEP}
  {\bfseries 09} (2012) 046} [\href{https://arxiv.org/abs/1203.3544}{{\ttfamily
  1203.3544}}].

\bibitem{Jensen:2012jh}
K.~Jensen, M.~Kaminski, P.~Kovtun, R.~Meyer, A.~Ritz and A.~Yarom,
  \emph{{Towards hydrodynamics without an entropy current}},
  \href{https://doi.org/10.1103/PhysRevLett.109.101601}{\emph{Phys. Rev. Lett.}
  {\bfseries 109} (2012) 101601}
  [\href{https://arxiv.org/abs/1203.3556}{{\ttfamily 1203.3556}}].

\bibitem{Hoyos:2014pba}
C.~Hoyos, \emph{{Hall viscosity, topological states and effective theories}},
  \href{https://doi.org/10.1142/S0217979214300072}{\emph{Int. J. Mod. Phys. B}
  {\bfseries 28} (2014) 1430007}
  [\href{https://arxiv.org/abs/1403.4739}{{\ttfamily 1403.4739}}].

\bibitem{AGD}
A.~A. Abrikosov, L.~P. Gorkov and I.~E. Dzyaloshinski, \emph{{Methods of
  Quantum Field Theory in Statistical Physics}}. Dover, 1963.

\bibitem{Fetter_Walecka}
A.~L. Fetter and J.~D. Walecka, \emph{{Quantum Theory of Many-Particle
  Systems}}. Dover, 2003.

\bibitem{Bruus_Frensberg}
H.~Bruss and K.~Flensberg, \emph{{Many-Body Quantum Theory in Condensed Matter
  Physics}}. Oxford, 2004.

\bibitem{sondheimer_wilson}
E.~H. Sondheimer and A.~H. Wilson, \emph{{The diamagnetism of free electrons}},
  \href{https://doi.org/10.1098/rspa.1951.0239}{\emph{Proc. R. Soc. London A}
  {\bfseries 210} (1951) 173}.

\bibitem{ErdelyiVol2}
H.~Bateman and A.~Erd\'elyi, \emph{{Higher Transcendental Functions, Vol. II}}.
  McGraw-Hill, 1957.

\bibitem{Lewin}
L.~Lewin, \emph{{Polylogarithms and associated functions}}. North Holland,
  1981.

\bibitem{Jaiswal:2015mxa}
A.~Jaiswal, B.~Friman and K.~Redlich, \emph{{Relativistic second-order
  dissipative hydrodynamics at finite chemical potential}},
  \href{https://doi.org/10.1016/j.physletb.2015.11.018}{\emph{Phys. Lett. B}
  {\bfseries 751} (2015) 548}
  [\href{https://arxiv.org/abs/1507.02849}{{\ttfamily 1507.02849}}].

\bibitem{Soo:1998ev}
C.~Soo, \emph{{Adler-Bell-Jackiw anomaly, the Nieh-Yan form, and vacuum
  polarization}}, \href{https://doi.org/10.1103/PhysRevD.59.045006}{\emph{Phys.
  Rev. D} {\bfseries 59} (1999) 045006}
  [\href{https://arxiv.org/abs/hep-th/9805090}{{\ttfamily hep-th/9805090}}].

\bibitem{Chandia:1997hu}
O.~Chand\'{\i}a and J.~Zanelli, \emph{{Topological invariants, instantons and
  chiral anomaly on spaces with torsion}},
  \href{https://doi.org/10.1103/PhysRevD.55.7580}{\emph{Phys. Rev. D}
  {\bfseries 55} (1997) 7580}
  [\href{https://arxiv.org/abs/hep-th/9702025}{{\ttfamily hep-th/9702025}}].

\bibitem{Rasulian:2023gtb}
I.~M. Rasulian and M.~Torabian, \emph{{On torsion contribution to chiral
  anomaly via Nieh\textendash{}Yan term}},
  \href{https://doi.org/10.1140/epjc/s10052-023-12331-y}{\emph{Eur. Phys. J. C}
  {\bfseries 83} (2023) 1165}
  [\href{https://arxiv.org/abs/2308.00578}{{\ttfamily 2308.00578}}].

\end{thebibliography}\endgroup


\end{document}